REVIEW ARTICLE **Open Access**

# Tuning the electronic states and superconductivity in alkali fulleride films

Ming-Qiang Ren[1*], Shu-Ze Wang[1], Sha Han[1], Can-Li Song[1,2*], Xu-Cun Ma[1,2*] and Qi-Kun Xue[1,2,3,4]

## Abstract

The successful preparation of superconducting alkali fulleride ($A_xC_{60}$, A = K, Rb, Cs) films using state-of-the-art molecular beam epitaxy overcomes the disadvantages of the air-sensitivity and phase separation in bulk $A_xC_{60}$, enabling for the first time a direct investigation of the superconductivity in alkali fullerides on the molecular scale. In this paper, we briefly review recent cryogenic scanning tunneling microscopy results of the structural, electronic, and superconducting properties of the fcc $A_xC_{60}$ films grown on graphitized SiC substrates. Robust *s*-wave superconductivity is revealed against the pseudogap, electronic correlation, non-magnetic impurities, and merohedral disorder. By controlling the alkali-metal species, film thickness, and electron doping, we systematically tune the $C_{60}{}^{x-}$ orientational orderings and superconductivity in $A_xC_{60}$ films and then complete a unified phase diagram of superconducting gap size vs electronic correlation and doping. These investigations are conclusive and elucidated that the *s*-wave superconductivity retains in alkali fullerides despite of the electronic correlation and presence of pseudogap.

## 1 Introduction

Alkali fullerides ($A_3C_{60}$, A = K, Rb, Cs) have been intensively studied since the first discovery of superconductivity in $K_3C_{60}$ in 1991 [1]. They exhibit a maximum superconducting transition temperature $T_c$ of ~ 40 K, which is the highest among the molecular superconductors and only gets surpassed by cuprates and iron-based superconductors [1–6]. Early studies found that the $T_c$ in $A_3C_{60}$ increases monotonically with increasing molecular spacing as $A^+$ cation size enlarges [5, 7–11]. The superconductivity in alkali fullerides seemly falls into the conventional Bardeen-Cooper-Schrieffer (BCS) regime where the intramolecular phonon plays a decisive role and the $T_c$ is mainly controlled by the density of states (DOS) at the Fermi level ($E_F$) [12, 13]. The full gap expected from a phonon mechanism scenario has been evidenced by various experimental techniques, such as scanning tunneling microscopy (STM) [14, 15] and nuclear magnetic resonance (NMR) [16].

From the view point of electronic structures, the superconductivity in $A_3C_{60}$ originates from the half-filled $t_{1u}$ band which possesses a narrow bandwidth ($W$ ~ 0.5 eV) because of the negligible charge transfers between $C_{60}$ molecules [7–9]. Though the on-site Coulomb repulsion ($U$ ~ 1.5 eV) significantly exceeds the $t_{1u}$ bandwidth [17, 18], metallicity is commonly observed in $K_3C_{60}$ and $Rb_3C_{60}$. The absence of a Mott localization was explained in terms of the $t_{1u}$ orbital degeneracy, which would strongly enhance the critical value $(U/W)_c$ of the Mott transition [19, 20].

Recent experiments on expanded $Cs_3C_{60}$ polymorphs reveal a dome-shaped phase diagram as a function of the $C_{60}{}^{3-}$ packing volume where superconductivity resides in proximity to a magnetic Mott-insulating state, highlighting the importance of electronic correlation in fulleride superconductors [21–26]. The measured Néel temperature $T_N$ is about 46 K in $Cs_3C_{60}$ with an orientational ordered A15 structure [22, 23], and is suppressed to 2.2 K in face-centered cubic (fcc) structured $Cs_3C_{60}$ due to the geometrical frustration [27–29].

* Correspondence: mqren@mail.tsinghua.edu.cn; clsong07@mail.tsinghua.edu.cn; xucunma@mail.tsinghua.edu.cn
[1]State Key Laboratory of Low-Dimensional Quantum Physics, Department of Physics, Tsinghua University, Beijing 100084, China
Full list of author information is available at the end of the article

In contrast to the atom-based cuprates high-$T_c$ superconductors, dynamic Jahn–Teller (JT) effect ($J_{JT}$) arising from the coupling of electrons to the on-molecule vibrations lifts the degeneracy of $t_{1u}$ orbital (see Fig. 1) and competes with the local exchange coupling ($J_H$) in fulleride superconductors [30, 31]. The larger energy of JT effect ($J_{JT} > J_H$) renders a low-spin state ($S = 1/2$) of $Cs_3C_{60}$ and an orbital disproportionation of filled electrons in fullerides [22, 24, 27, 32–35]. From this point of view, $Cs_3C_{60}$ (fcc and A15) polymorphs have been classified as magnetic Mott–Jahn–Teller insulators (MJTIs) [8, 10, 21–26]. The cooperation between strong correlation ($U/W$) and JT distortion stabilizes the magnetism. With increasing pressure or bandwidth $W$, $Cs_3C_{60}$ involves from an MJTI phase into a JT metal (metallic phase with JT effect), accompanied with a dome-shaped superconductivity upon cooling [24]. The superconductivity in $A_3C_{60}$ is expected to be modulated by these diversified phases. For example, recent studies found that the optimized $T_c$ in fcc $Rb_xCs_{3-x}C_{60}$ occurs at the boundary between a JT metal and a conventional metal [24], and the reduced gap ratio $2\Delta/k_BT_c$ increases dramatically in the expanded regime where JT effect matters [24, 36].

As discussed above, the competition/cooperation between bandwidth, on-molecule electron-phonon coupling, and electronic correlation rises much complexity in $A_3C_{60}$ and obstacles a comprehensive understanding on the electronic structure and superconductivity. In particular, the complexity arising from the air sensitivity and phase separation of $A_3C_{60}$ polymorphs, limits the experimental techniques primarily to NMR and magnetization measurements [10, 21–28]. Local measuring probes, such as STM, may circumvent the above issues but thus far have been limited to several mechanically cleaved $A_3C_{60}$ crystals under the atmospheric or argon environment [14, 15, 37–39] and a few non-superconducting $K_xC_{60}$ films on metal substrates [40–42]. Preceding STM experiments on cleaved $K_3C_{60}$, $Rb_3C_{60}$, and $Rb_2CsC_{60}$ reveal a highly inhomogeneous superconducting gap, because of the strongly varied stoichiometry. The poor quality of the cleaved $A_3C_{60}$ surfaces hinders the atomically resolved imaging and further measurements [14, 15, 37–39].

Until now, there remains a couple of important issues in $A_3C_{60}$ superconductors. First, the microscopic mechanism of fulleride superconductivity, with the conventional phonon-mediated pairing or unconventional electronic pairing or a synergy between them, remains controversial [17, 24, 32, 43]. Besides, the interplay between electronic correlation and molecular JT instability might reduce the dimensionality of low-lying states from 3D to 2D [30, 34]. A fundamental question thus arises as to how the reduced dimensionality affects the superconductivity in fullerides. What is more, the phase diagram in $A_3C_{60}$ is commonly documented as a function of pressure or $C_{60}^{3-}$ packing volume, in contrast to the doping-controlled phase diagram in cuprates. Therefore, the successful preparation of high-quality superconducting $A_3C_{60}$ films with precisely controlled film thickness and electron doping is crucial to clarify these issues [44–46].

This article aims to give a brief introduction to recent STM results on the fcc $A_xC_{60}$ superconducting films and is organized as follows. First, we describe the diversified orientational orderings of $C_{60}^{x-}$ in $A_xC_{60}$ films. Then, we proceed to discuss the $s$-wave superconductivity and its response to impurities or merohedral disorders. After this, we turn to the tunability of electronic structure and superconductivity in $A_xC_{60}$ films via controlling the electron doping $x$ and the film thickness. Finally, a doping-controlled phase diagram of $A_xC_{60}$ is summed up, which gives strong restrictions on the superconducting mechanism in alkali fullerides.

## 2 State-of-the-art MBE growth

Bulk $A_xC_{60}$ crystalizes into either a fcc structure (see Fig. 2a) or an orientational ordered A15 structure ($Cs_3C_{60}$) in which $C_{60}$ locates on the body-centered-cubic (bcc) lattice [21, 23, 27, 47, 48]. The preparation of $A_xC_{60}$ films (see Fig. 2b) using molecular beam epitaxy (MBE) includes two steps [44–46]. Firstly, $C_{60}$ molecules were evaporated onto the bilayer-graphene dominated SiC surfaces with well-controlled film thickness. Secondly, alkali-metal atoms were deposited onto the pristine $C_{60}$ films followed by a moderate annealing. The as-grown $A_xC_{60}$ films crystallize into a (111)-orientated fcc structure, as depicted in Fig. 2c. It is clear that the $C_{60}$ molecules assemble into a hexagonal lattice. Notably, $C_{60}$ molecules in $n$ ML and ($n$ + 2) ML occupy two symmetry-inequivalent sites in the unit cell of the middle layer, suggesting an ABC stacking as expected from the fcc structure [45]. On the other hand, the in-plane and inter-layer spacings of $C_{60}$ molecules in $K_3C_{60}$ films is measured to be 8.1 ± 0.1 Å and 10.0 ± 0.1 Å, consistent well with that of the fcc-structured bulk $K_3C_{60}$ crystals.

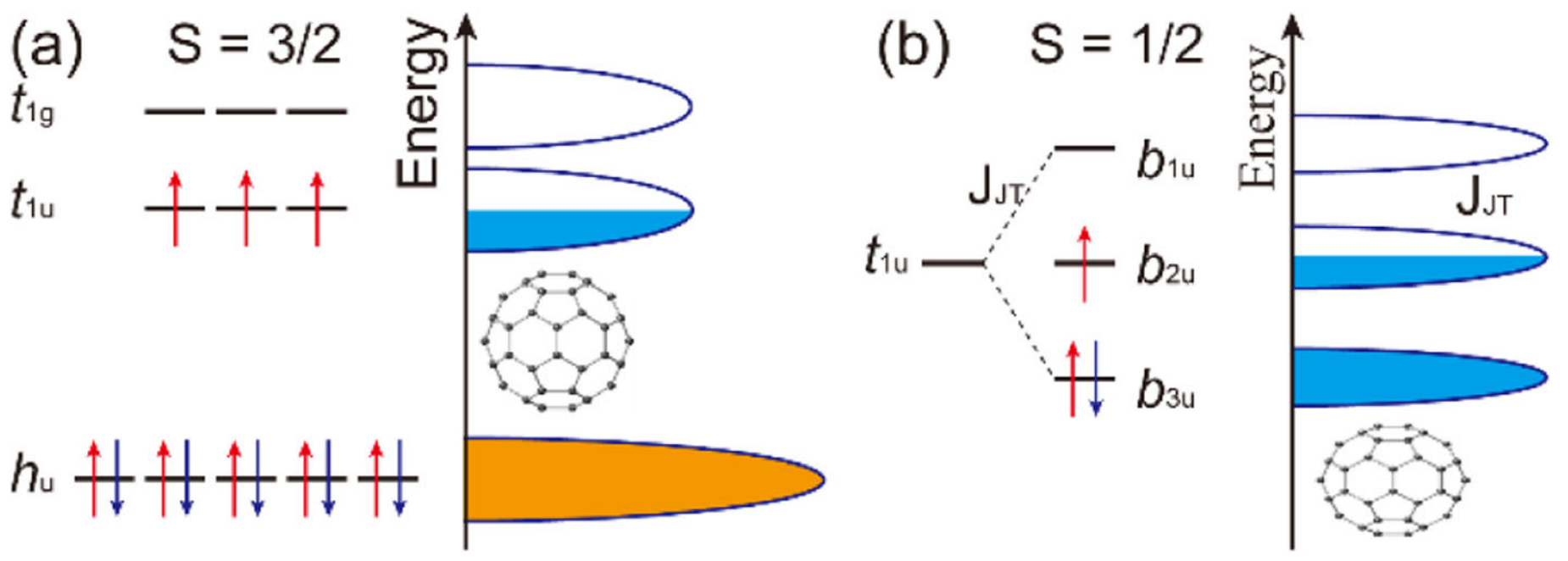

**Fig. 1** Sketches of molecular orbital energy of $C_{60}^{3-}$ anions and simplified electronic structure of solid $A_3C_{60}$ (**a**) with and (**b**) without JT effect. The JT distortion splits the triple degenerated $t_{1u}$ orbital into $b_{1u}$, $b_{2u}$, and $b_{3u}$ orbitals

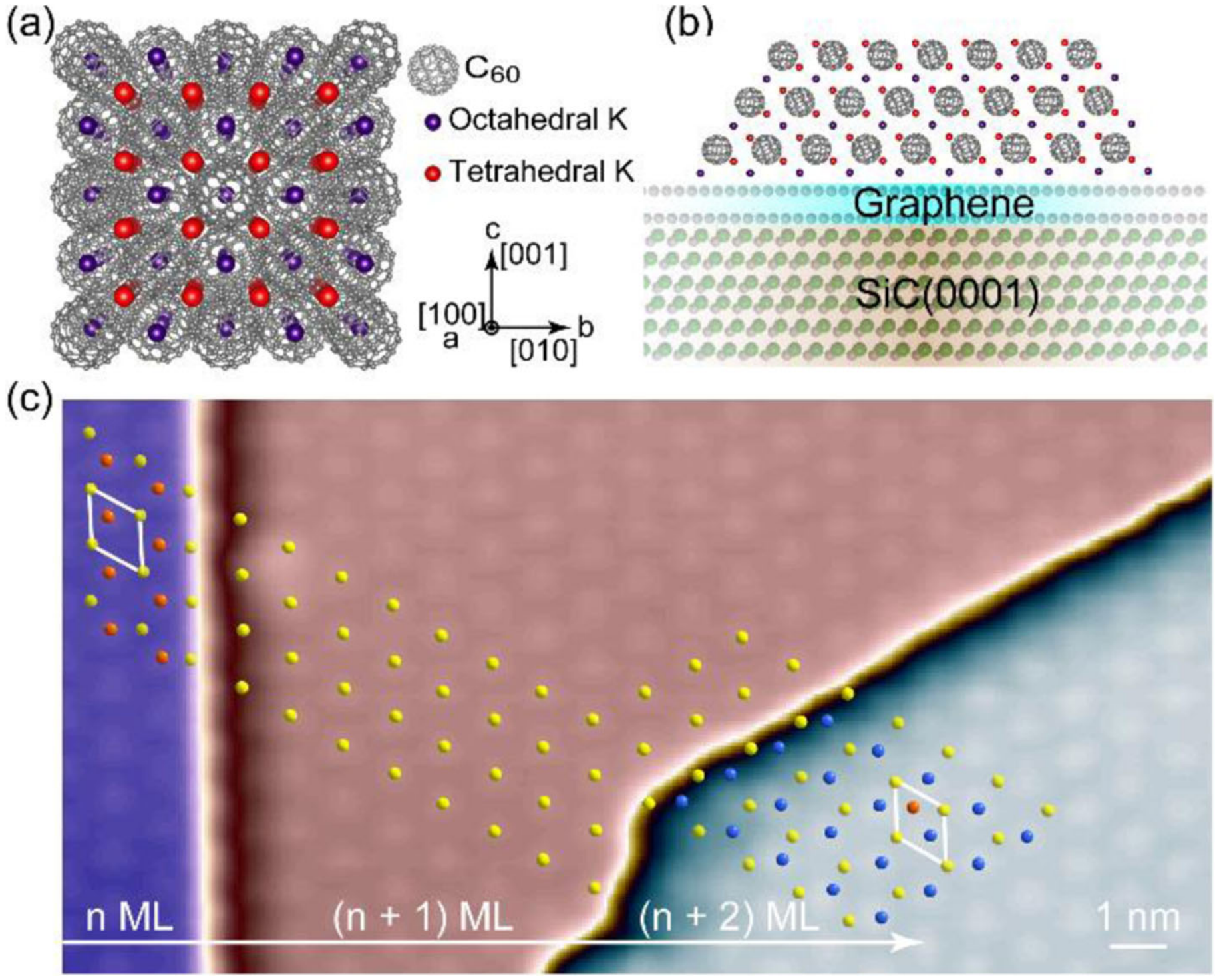

**Fig. 2** (**a**) Crystal structure of fcc $K_3C_{60}$. (**b**) Schematic side view of $K_3C_{60}$(111) films on graphitized SiC substrate. (**c**) STM topography ($V$ = 3.0 V, $I$ = 10 pA) showing three consecutive $K_3C_{60}$ layers. Taken from ref. [45]

## 3 Orientational orderings

Unlike the atom-based superconductors, molecule-based alkali fullerides exhibit additional degrees of freedom related to the molecular orientation. The merohedral disorder, one typical arrangement of $C_{60}$ orientations, has been confirmed in $A_3C_{60}$ [47, 48]. However, the role of the merohedral disorder has been neglected in most theoretical calculations, though some calculations show that it would significantly alter the low-energy electronic structures [49, 50]. On the other hand, experimental evidences for its impact on superconductivity and the electronic structures are even rarer. Thus, it becomes increasingly important to visualize and quantify the merohedral disorder and $C_{60}$ orientational orderings in $A_3C_{60}$.

### 3.1 Undoped $C_{60}$ films

STM topographies of $C_{60}$ films on graphene/SiC reveal five distinct $C_{60}$ orientations, namely, $C_{60}$ molecules orientate a hexagon (H), a pentagon (P), a 6:6 bond (H:H), a 6:5 bond (H:P), or a carbon apex (CA) pointing up (see Fig. 3a, b) [44]. It is worth mentioning that the H-orientated $C_{60}$ molecules exhibit a tri-star-like topography. This is because that C-C bonds connecting a pentagon (P) and a hexagon (H) have relatively lower electron density and thus the $C_{60}$ pentagons would brighten in the empty-states, as colored in red in Fig. 3c. Besides, the $C_{60}$ orientations in both monolayer and bilayer films arrange into a quasi-(2 × 2) superstructure with a short-range correlation (white rhombus in Fig. 3a, b). The submolecular structures and the 2 × 2 superstructure have been routinely observed in $C_{60}$ films grown on either the graphene substrate or metal substrates, and will not be discussed here [52–55].

### 3.2 Merohedral disorder in $A_3C_{60}$ films

With alkali-metal atoms intercalated into the tetrahedral and octahedral sites in fcc $A_3C_{60}$, the repulsive A-$C_{60}$

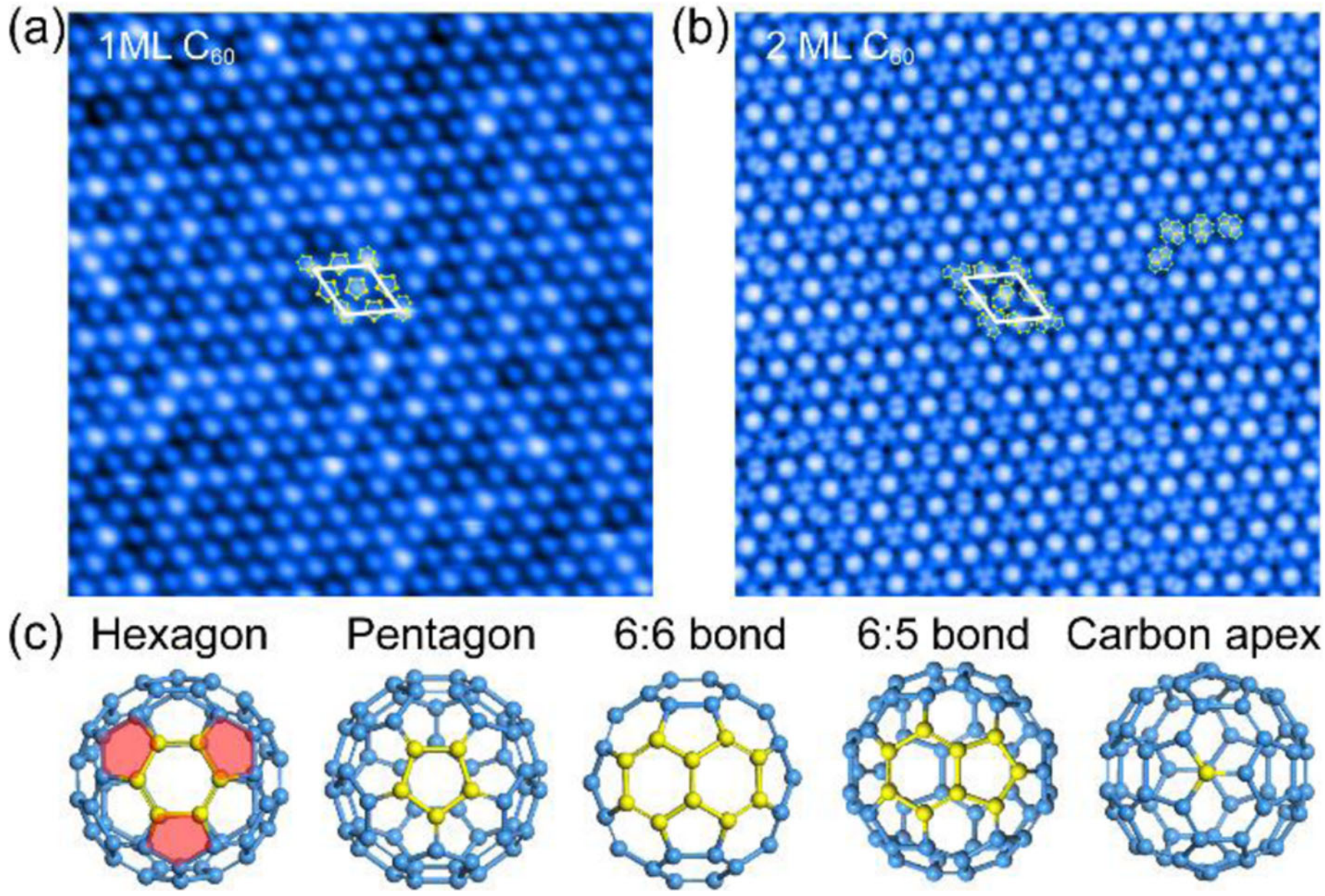

**Fig. 3** $C_{60}$ orientations in undoped $C_{60}$ films. (**a**, **b**) Empty-state STM images (10 × 10 $nm^2$) of monolayer and bilayer $C_{60}$ films. The white rhombuses mark the unit cells of the quasi-(2 × 2) superstructures. (**c**) Sketch maps for five orientations (as labeled) of $C_{60}$ molecules viewed from [51]. Taken from Ref. [44]

interaction strongly favors two standard orientations (see Fig. 4a), which are related by 90° rotation about [56] direction or 44.48° rotation about [51, 57]. Both of them are H-orientated when viewed from [51]. The merohedral disorder thus refers to the random occupation of the two standard orientations by $C_{60}$ molecules at low temperature.

Experimentally, the $C_{60}$ molecules in $A_3C_{60}$ films grown on graphene/SiC are universally H-orientated, but their orderings are strongly fluctuated by the film thickness and alkali-metal species [44–46]. The STM topographies of $A_3C_{60}$ thick films (~ 9 ML) are shown in Fig. 4d-f. The H-orientated $C_{60}$ molecules with a tri-star-like feature assemble into an hexagonal 1 × 1 lattice. Notably, $C_{60}$ molecules in $K_3C_{60}$ behave the same orientation with a long-range orientational correlation (no merohedral disorder). However, the long-range ordering melts in $Rb_3C_{60}$ and nanoscale patches occur as colored in Fig. 4d, indicating a moderate merohedral disorder. Finally, the orientational correlation completely disappears in $Cs_3C_{60}$ (strong merohedral disorder). In order to quantify the merohedral disorder, we calculated the averaged orientational correlation function $<\cos(\theta_{ij})>$ [58], in which $\theta_{ij} = \theta_i - \theta_j$ denotes the angle between nearest neighbor $C_{60}$ molecules. As shown in Fig. 4b, the value of $<\cos(\theta_{ij})>$ equals to 1 in $K_3C_{60}$ films, as expected from the complete orientational ordering [44–46]. With increasing $A^+$ cations size, $<\cos(\theta_{ij})>$ decreases dramatically, indicating an enhanced merohedral disorder. This change is probably due to the weakening repulsions between neighboring $C_{60}{}^{3-}$ as a result of lattice expansion [59]. Despite of the significant change of $<\cos(\theta_{ij})>$ with alkali-metal atoms, $<\cos(\theta_{ij})>$ is found to be almost independent of the electron doping $x$, if the deviation $x - 3$ is not very large (see Fig. 4c), indicating an essentially unchanged merohedral disorder against $x$. The impact of merohedral disorder on the electronic structures and superconductivity will be discussed in section 4.3 and section 5, respectively.

Not only the $A^+$ cations, $C_{60}$ orientational orderings in $A_3C_{60}$ are also associated with the film thickness, which holds particularly true for monolayer films. Monolayer $K_3C_{60}$ and $Cs_3C_{60}$ exhibit a √3 × √3 superstructure of the H-orientated $C_{60}$ (see Fig. 4g, h) [44, 45], suggesting a non-negligible graphene-$C_{60}$ interaction. For $Cs_3C_{60}$, the √3 × √3 superstructure exists solely to the monolayer and is absent in thicker (> 2 ML) films [44]. For $K_3C_{60}$, a weak √3 × √3 superstructure persists in the bilayer but disappears completely in a trilayer film [45]. The same √3 × √3 superstructure has also been observed in monolayer $K_3C_{60}$ grown on Au (111), and is attributed to the K-induced reconstruction therein [40–42].

### 3.3 Orientational orderings in $Cs_xC_{60}$ ($x$ = 1, 2, 4) films

As discussed above, the $C_{60}$ molecules in $A_3C_{60}$ are all H-orientated, which is stabilized by the repulsive A-$C_{60}$ interaction. Thus, dramatic changes of $C_{60}$ orientations are expected to occur in $A_xC_{60}$ ($x$ = 1, 2, 4), since the occupation sites of alkali-metal atoms are different [60]. In $A_3C_{60}$, alkali-metal atoms occupy all the octahedral and tetrahedral sites. In $A_1C_{60}$ ($A_2C_{60}$), only the octahedral sites (two tetrahedral sites) are occupied. For $A_4C_{60}$ films, the fourth alkali-metal atom occupies the octahedral site for the topmost $C_{60}$ layer [45]. This differs to that in $Li_4C_{60}$ bulk where the fourth Li occupies the 32f sites [61]. We have intensively studied the orientational orderings in $Cs_xC_{60}$ ($x$ = 1, 3 ,4) [44], and the results are summarized in Fig. 5. In bilayer $Cs_xC_{60}$, which is weakly correlated with the graphene substrates, the $C_{60}$ orientations change from CA, H:P, H to H:H with $x$ = 1, 2, 3, 4, respectively. However, $C_{60}$ molecules in monolayer $Cs_1C_{60}$ and $Cs_2C_{60}$ all become H-orientated, confirming again the role of substrate to $C_{60}$ orientations. Besides, the periodicity of superstructures in the monolayer $Cs_xC_{60}$ is found to strongly rely on the electron doping $x$.

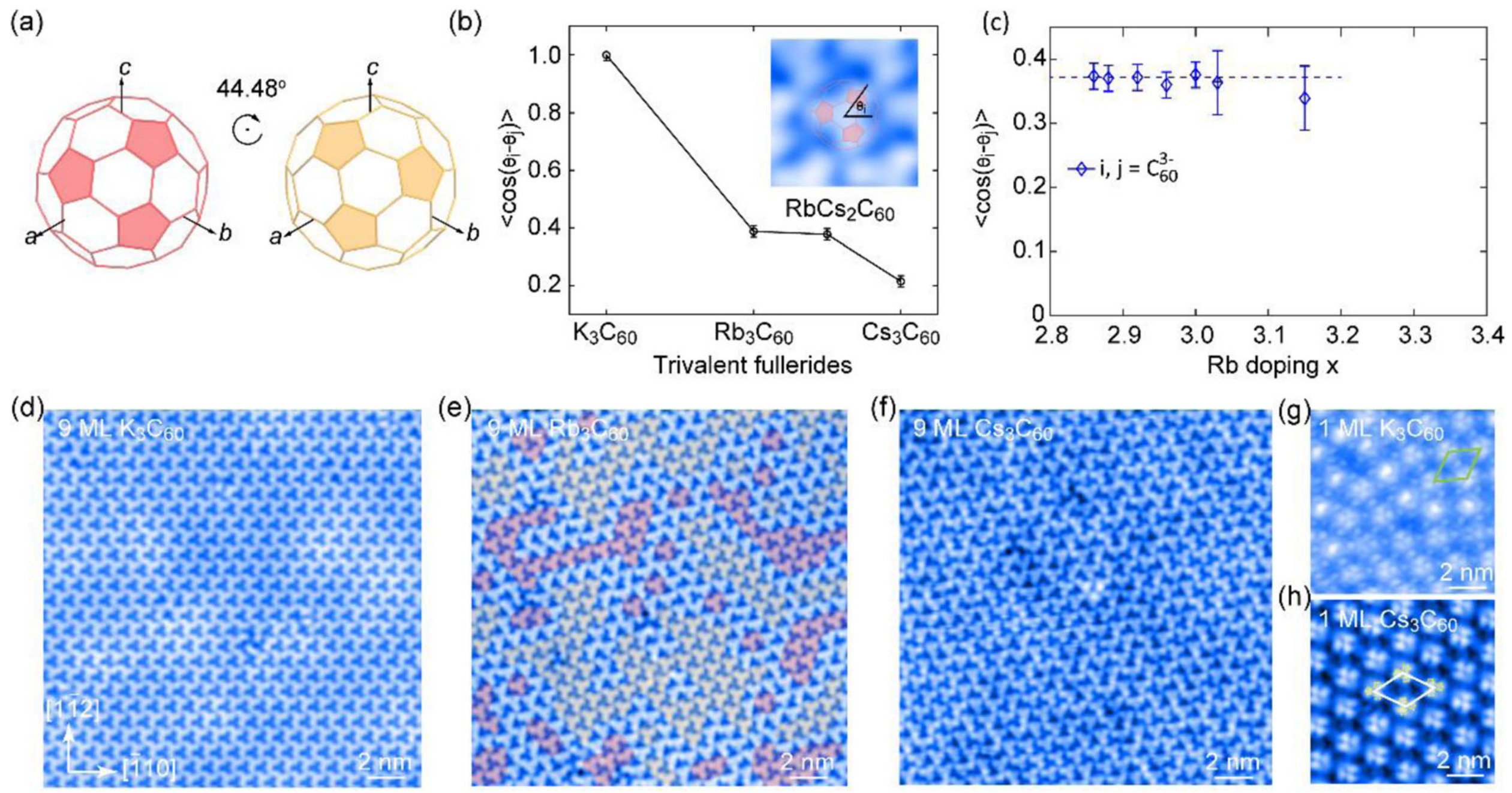

**Fig. 4** Orientational orderings in $A_3C_{60}$ films. (**a**) Two standard $C_{60}$ orientations viewed along the [51] direction. (**b**) Orientational correlation in 9 ML $A_3C_{60}$ films. Inset shows the definition of $\theta_i$. (**c**) Averaged orientational correlation in 9 ML $Rb_xC_{60}$ films as a function of doping $x$. (**d**-**f**) STM topographies of 9 ML $A_3C_{60}$ films. (**g**, **h**) STM topographies of monolayer $K_3C_{60}$ and $Cs_3C_{60}$ films revealing a $\sqrt{3} \times \sqrt{3}$ superstructure. Taken from ref. [44–46]

In summary, $A_xC_{60}$ films grown on graphene/SiC exhibit diversified $C_{60}$ orientations and their orderings, which are closely associated with the alkali-metal atoms (K, Rb, Cs), electron doping $x$, and film thickness, as listed in Table 1. Electron doping $x$ dominates the $C_{60}$ orientations in bilayer or thicker $A_xC_{60}$ films as well as the periodicity of superstructures in $A_xC_{60}$ monolayers. Film thickness dramatically impacts on the $C_{60}$ orientations in monolayer films, because of the non-negligible graphene-$C_{60}$ interaction. The alkali-metal atoms mainly control the merohedral disorder of the two standard H-orientated $C_{60}$ molecules in $A_3C_{60}$.

## 4 Low-energy electronic states and superconductivity

### 4.1 Decoupling effect in undoped $C_{60}$ films

Before starting the $A_3C_{60}$ films, it is necessary to introduce some important results on the undoped $C_{60}$ films. Monolayer $C_{60}$ grown on graphene is an insulator with a HOMO-LUMO gap of 3.5 eV, consistent fairly well with that in solid $C_{60}$ [62], but is relatively larger than that grown on metal surfaces [63–65]. This indicates a minor charge transfer and substrate-induced screening from underlying graphene compared to metal substrates [52, 66]. This decoupling effect enables it readily to deduce the intrinsic electronic properties of $C_{60}$ films and $A_3C_{60}$ overlayers [44], though graphene substrate plays a non-negligible role in determining the $C_{60}$ orientations in monolayers. Notably, the triply degenerate $t_{1u}$ orbital has been lifted into two discrete peaks (see Fig. 6), indicating the presence of JT distortion in charged fullerenes [67, 68], even though the charge transfer is very small.

### 4.2 Insulating ground states in $Cs_xC_{60}$

With alkali-metal intercalated, the transferred electrons will occupy the triply degenerate $t_{1u}$ orbital. The charge transfer between alkali-metal atoms and $C_{60}$ molecules is considered to be essentially complete [69]; thus, $x$ also indicates the electron doping for per $C_{60}{}^{x-}$. Considering a simple band theory, metallicity is expected in $A_xC_{60}$ if $0 < x < 6$, since $t_{1u}$ orbital is partially filled. Experimentally, all measured $Cs_3C_{60}$ films (monolayer, bilayer, and 9 ML) grown on graphene substrates are insulating, characteristic of apparent energy gaps around $E_F$ [44, 46]. Moreover, monolayer and bilayer $Cs_xC_{60}$ with other electron doping $x$ = 1, 2, 4 are also insulators [44]. The case for $Cs_2C_{60}$ and $Cs_3C_{60}$ monolayers is depicted in Fig. 7a.

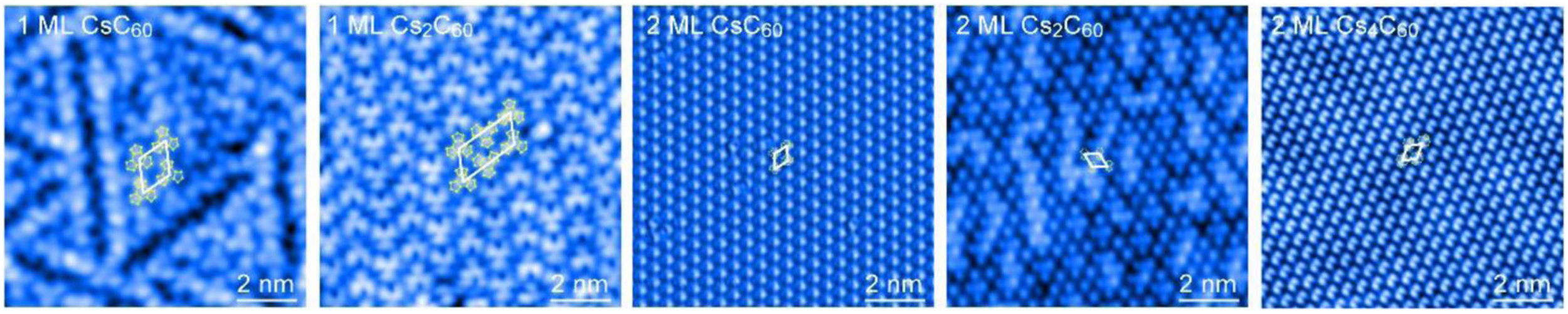

**Fig. 5** STM topographies of 1 ML $Cs_xC_{60}$ ($x$ = 1, 2) and 2 ML $Cs_xC_{60}$ ($x$ = 1, 2, 4) as labeled. The $C_{60}$ orientations and the unit cells are marked for clarity. Taken from ref. [44]

The robust insulating behavior speaks highly of the role of both on-molecule Coulomb repulsion and JT instability in determining the electronic states in fulleride films [8, 24, 26]. The lift of the triply degenerate $t_{1u}$ orbitals via a JT coupling is evidenced by the accompanied satellite peaks (up arrows in Fig. 7a) beyond the gaps [44]. On the basis of JT coupling and correlation $U$, it is straightforward to explain the insulating ground states of $Cs_xC_{60}$, as illustrated in Fig. 7b. For even $x$ = 2 and 4, the strong JT effect splits the $t_{1u}$ orbital into sub-bands ($b_{1u}$, $b_{2u}$, $b_{3u}$) and results in a charge-disproportionated insulator [33–35], whereas for odd $x$ = 1 and 3, the strong correlation $U$ further splits the partially filled sub-band into upper and lower Hubbard bands. Thus, the two nearest DOS peaks/kinks next to $E_F$ in the d$I$/d$V$ spectra, as marked by the triangles in Fig. 7a, can be safely assigned to the JT subbands ($x$ = 2, 4) or UHB/LHB ($x$ = 1, 3).

### 4.3 Emergent superconductivity in $K_3C_{60}$ and $Rb_3C_{60}$

In $A_3C_{60}$ crystals, the cation size of alkali-metals controls the $C_{60}{}^{3-}$ packing volume and dramatically affects the bandwidth, and thus the superconductivity and low-energy electronics states. Similar phenomenon is also observed in the $A_3C_{60}$ films grown on graphene substrates, and the case for 9 ML films is shown in Fig. 8a. In the most expanded $Cs_3C_{60}$ films, the strong electronic correlation $U/W$ and JT effect lead to the opening of a Mott insulating gap. Thus, 9 ML $Cs_3C_{60}$ film is a MJTI [44]. With reduced $A^+$ cation size or lattice constant, electron delocalizes as $U/W$ decreases, and metallicity as well as a superconductivity is realized in 9 ML $K_3C_{60}$ and $Rb_3C_{60}$ films. The evolution of the superconducting gap size $\Delta$ and $T_c$ with alkali-metal atoms is summarized in Fig. 8b, c. Similar to that in pressurized $Cs_3C_{60}$ [21, 22], a dome-shaped dependence of both the $\Delta$ and $T_c$ is clearly revealed, though only discrete values of lattice constant are accessible by controlling the alkali-metal atoms. Moreover, $T_c$ of 9 ML $K_3C_{60}$ and $Rb_3C_{60}$ is almost identical to that of their bulk counterparts [1, 2], suggesting a similarity between 9 ML $A_3C_{60}$ films and the bulk crystals. The lower $T_c$ in a $RbCs_2C_{60}$ films compared to its bulk form may result from an inhomogeneous distribution of Rb, Cs atoms, which may alter the actual chemical composition of the topmost layer.

**Table 1** A brief summary of $C_{60}$ orientations and their orderings in $A_xC_{60}$ films prepared on graphene. H, CA, H:P, H:H donate the $C_{60}$ orientations. OD is short for merohedral disorder. Taken from ref. [44–46]

| OD | $Cs_1C_{60}$ | $Cs_2C_{60}$ | $K_3C_{60}$ | $Rb_3C_{60}$ | $Cs_3C_{60}$ | $K_4C_{60}$ | $Cs_4C_{60}$ |
|---|---|---|---|---|---|---|---|
| 1 ML | H | H | H | | H | | |
| | (1 × 1) | (1 × 2) | (√3 × √3) | | (√3 × √3) | | |
| 2 ML | CA | H:P | H | H | H | H:H | H:H |
| | (1 × 1) | (1 × 1) | (Weak √3 × √3) | Moderate OD | Strong OD | (1 × 1) | (1 × 1) |
| 3 ML | | | H | H | | H:H | |
| | | | Without OD | Moderate OD | | (1 × 1) | |
| 9 ML | | | H | H | H | H:H | |
| | | | Without OD | Moderate OD | Strong OD | (1 × 1) | |

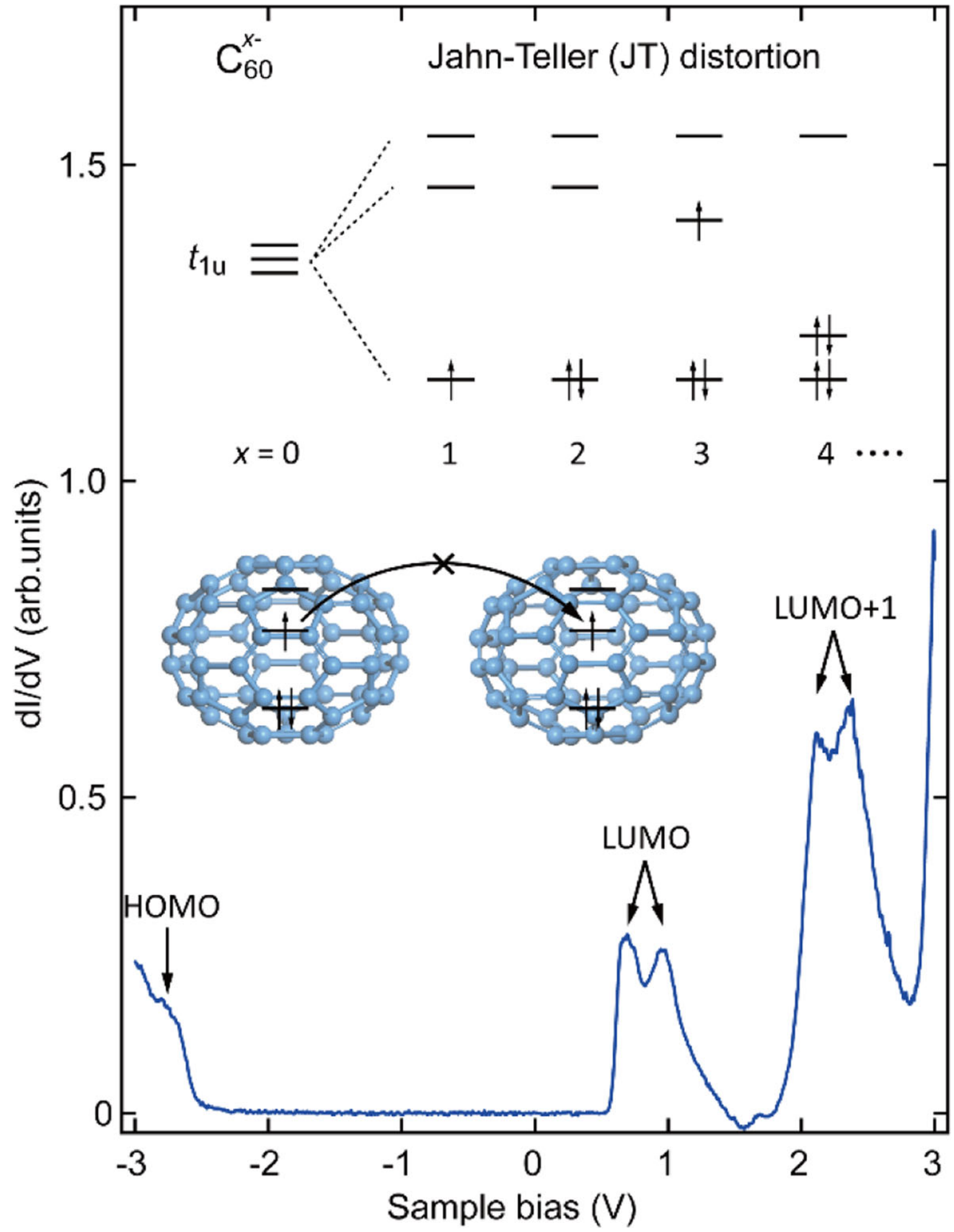

**Fig. 6** d*I*/d*V* spectrum of 1 ML $C_{60}$ film grown on graphene. The inset shows the modified molecular orbitals (top) of $C_{60}{}^{x-}$ anions by JT distortion (bottom). Taken from ref. [44]

Despite of the bandwidth-controlled superconductor-Mott-insulator transition (SMIT), the d$I$/d$V$ spectra also reveal two important aspects. Firstly, 9 ML $K_3C_{60}$ exhibits two sharp peaks around −0.4 eV and 0.1 eV, following the calculated electronic structure of $A_3C_{60}$ assuming a fixed orientation [49, 70]. The ordered arrangement of $C_{60}$ molecules in 9 ML $K_3C_{60}$ has been evidenced in Fig. 4d. However, the sharp peaks are erased in 9 ML $Rb_3C_{60}$ and $RbCs_2C_{60}$ with enhanced merohedral disorder, leading to a smooth variation of electronic DOS, as theoretically anticipated [49]. Secondly, the estimated bandwidth of $t_{1u}$ orbital, defined as the spacing between the two conductance minima (dashed lines in Fig. 8a) below and above $E_F$, is significantly larger than the calculated value of ~ 0.5 eV [49, 70]. Such a discrepancy should originate from the JT instability and Coulomb interaction omitted by three-band first principle calculations [49, 70]. If the JT-induced sub-band splitting and electronic correlation are involved, calculations give a substantially increased $t_{1u}$ bandwidth, as observed here [26, 32, 71].

### 4.4 *s*-wave pairing

We declare that a fully gapped superconductivity as well as a pseudogap emerges robustly in $K_3C_{60}$ and $Rb_3C_{60}$ films (> 3 ML), whereas the monolayer and bilayer films are not superconducting. The superconducting gap in a $K_3C_{60}$ trilayer is exemplified in Fig. 9a, c. Despite some heterogeneity in the coherence peaks, the d$I$/d$V$ spectra reveal completely vanished DOS near $E_F$. At some positions, the gap exhibits pronounced coherence peaks and could be well fitted with a single BCS-type isotropic $s$-wave gap function [72]. The averaged Δ, defined as half the distance between the two coherence peaks, is estimated to be ~5.7 meV. Figure 9 plots the temperature dependence of the superconducting gap with elevated temperature. Apparently, the superconducting gap is gradually suppressed, but evolves

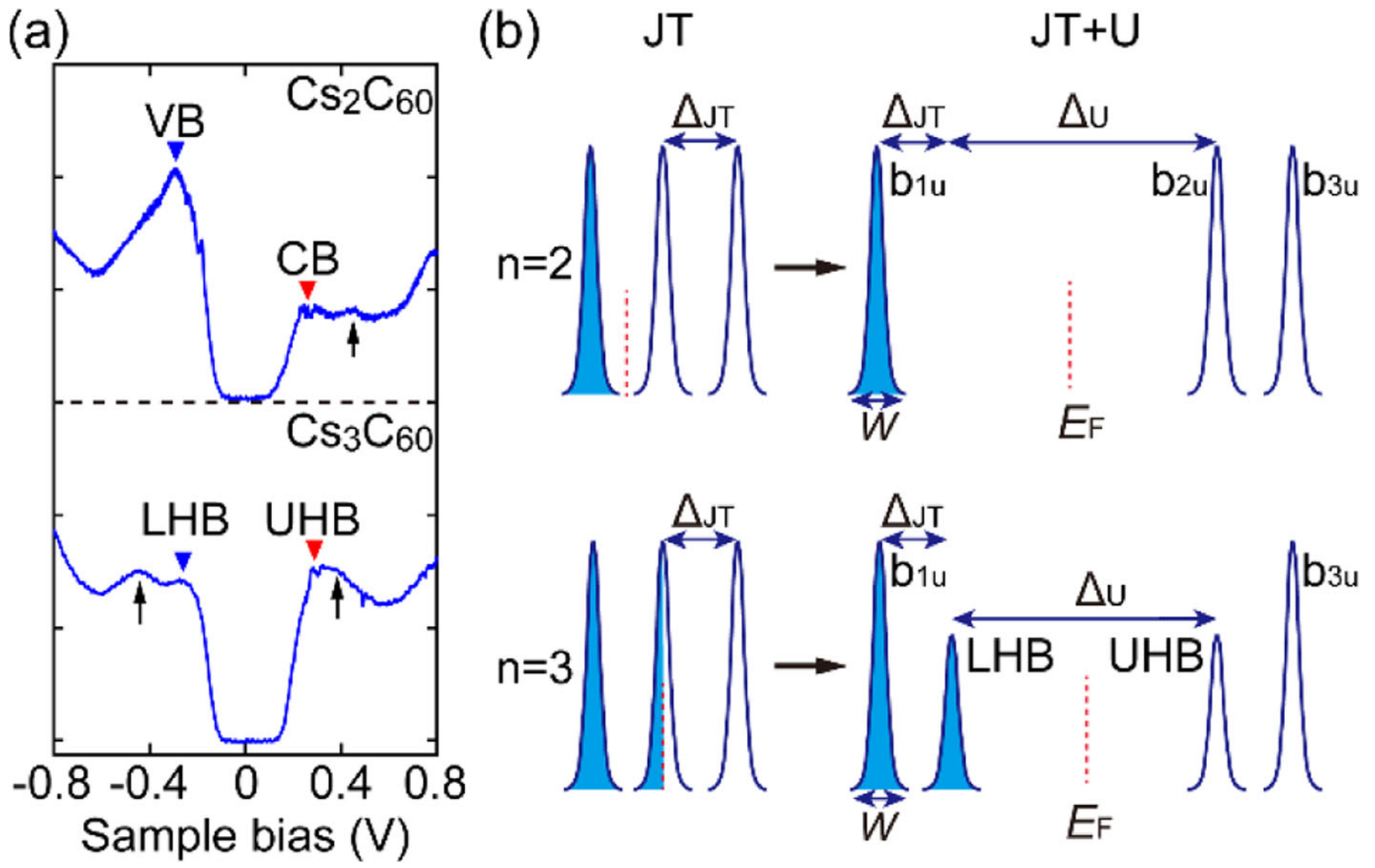

**Fig. 7** (**a**) d*I*/d*V* spectra of 1 ML $Cs_2C_{60}$ and $Cs_3C_{60}$. The triangles and up arrows donate the LHB/UHB or VB/CB and the satellite peaks from JT distortion, respectively. Taken from ref. [44]. (**b**) Schematic diagram of the universal insulating states in $Cs_xC_{60}$

continuously into a normal state quasiparticle gap rather than vanishes above $T_c$. This is reminiscent of the pseudogap phenomena in the underdoped cuprates [73, 74]. The enhanced $T_c$ in $K_3C_{60}$ trilayer (~ 22 K) compared to 9 ML films may result from an increased correlation with reduced film thickness, as will discussed in section 5.

The *s*-wave superconductivity is also confirmed by the observation of the isotropic vortices under magnetic fields (see Fig. 10a) [45, 46] By fitting the radial dependence of the normalized zero-energy conductance, we deduce the coherence length ξ and the angle dependence of ξ in the case of $K_3C_{60}$ trilayer is plotted in Fig. 10c. Apparently, ξ is angle independent, consistent with the isotropic *s*-wave symmetry, since $\xi \propto 1/\Delta$ in BCS theory. Besides, ξ is measured to be 2.6 nm in $K_3C_{60}$ and 1.5 nm in $Rb_3C_{60}$ films, in accord with that in their bulk counterparts [45, 46, 75]. Such a small ξ agrees with a local pairing picture, where intramolecular JT phonons play an important role in the formation of cooper pairs [11, 26, 76]. Furthermore, the d$I$/d$V$ spectra across the vortex (see Fig. 10e) reveal again the pseudogap feature. In contrast to cuprates, no spatial charge density modulation of the normal state quasiparticles is observed inside the vortex (see Fig. 10b), indicating that charge density wave (CDW) correlations are not responsible for the opening of the pseudogap [74]. A close examination reveals that the pseudogap is suppressed in 9 ML $K_3C_{60}$ than that in 3 ML $K_3C_{60}$ (see Fig. 10d). Previously, the pseudogap was not observed in bulk fullerides [15]. This implies that pseudogap might be a general phenomenology of 2D superconductors [77].

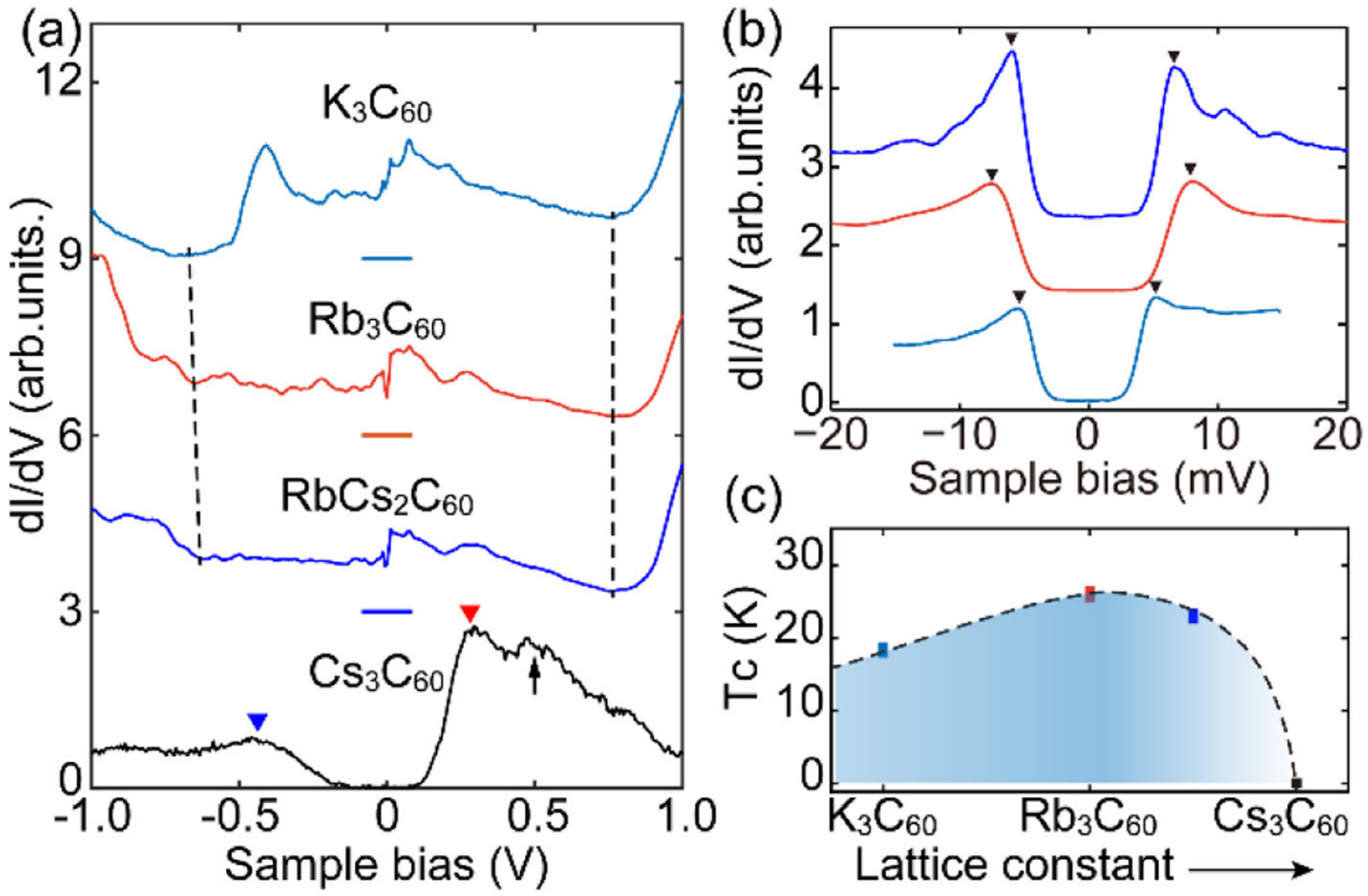

**Fig. 8** Bandwidth control of 9 ML $A_3C_{60}$. (**a**) Low energy-scale d*I*/d*V* spectra. The triangles and up arrows donate the LHB/UHB and the satellite peaks from JT distortion, respectively. (**b**) Evolution of superconducting gap taken at 4.5 K. The spectra are color-coded to that in (**a**). (**c**) Electronic phase diagram of 9 ML $A_3C_{60}$ films. Taken from ref. [45, 46]

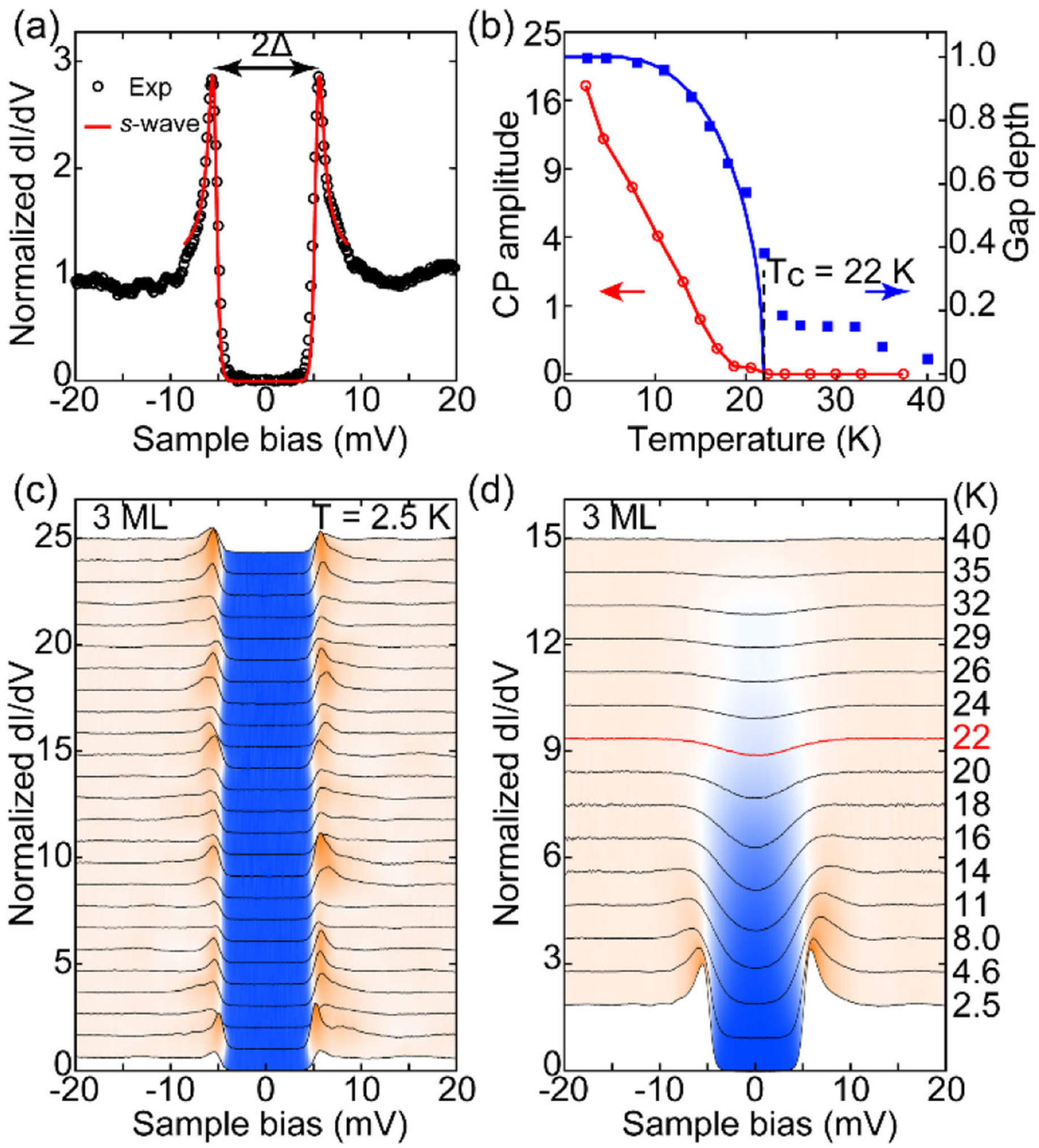

**Fig. 9** Superconductivity in 3 ML $K_3C_{60}$. (**a**) d*I*/d*V* spectrum taken at 2.5 K and its best fit to an isotropic *s*-wave superconducting gap. (**b**) Temperature dependence of the coherence peak (CP) amplitude and gap depth, revealing a $T_c$ of 22 K. (**c**) Grid d*I*/d*V* spectra taken at 2.5 K in a field of view of 20 nm × 20 nm. (**d**) Spatially averaged d*I*/d*V* spectra as a function of temperature. Taken from ref. [45]

In 9 ML $Rb_3C_{60}$ and 9 ML $RbCs_2C_{60}$, the superconducting gap shows a spatial inhomogeneity in both the coherence peak amplitude and the gap size Δ [46]. Such inhomogeneity is proved to be irrelevant to the local merohedral disorder, since apparent coherence peaks can also exist on the regions between nearest neighbor merohedral domains [46]. Moreover, a careful examination of the superconducting $Rb_3C_{60}$ films reveals that the coherence peak amplitude scales inversely with the Δ [46]. This is unexpected by the conventional wisdom of BCS picture, and may be ascribed to a coexistence of competing order, such as the observed pseudogap phase.

To further examine the *s*-wave pairing, we probe the local quasiparticle states around impurities or disorders,

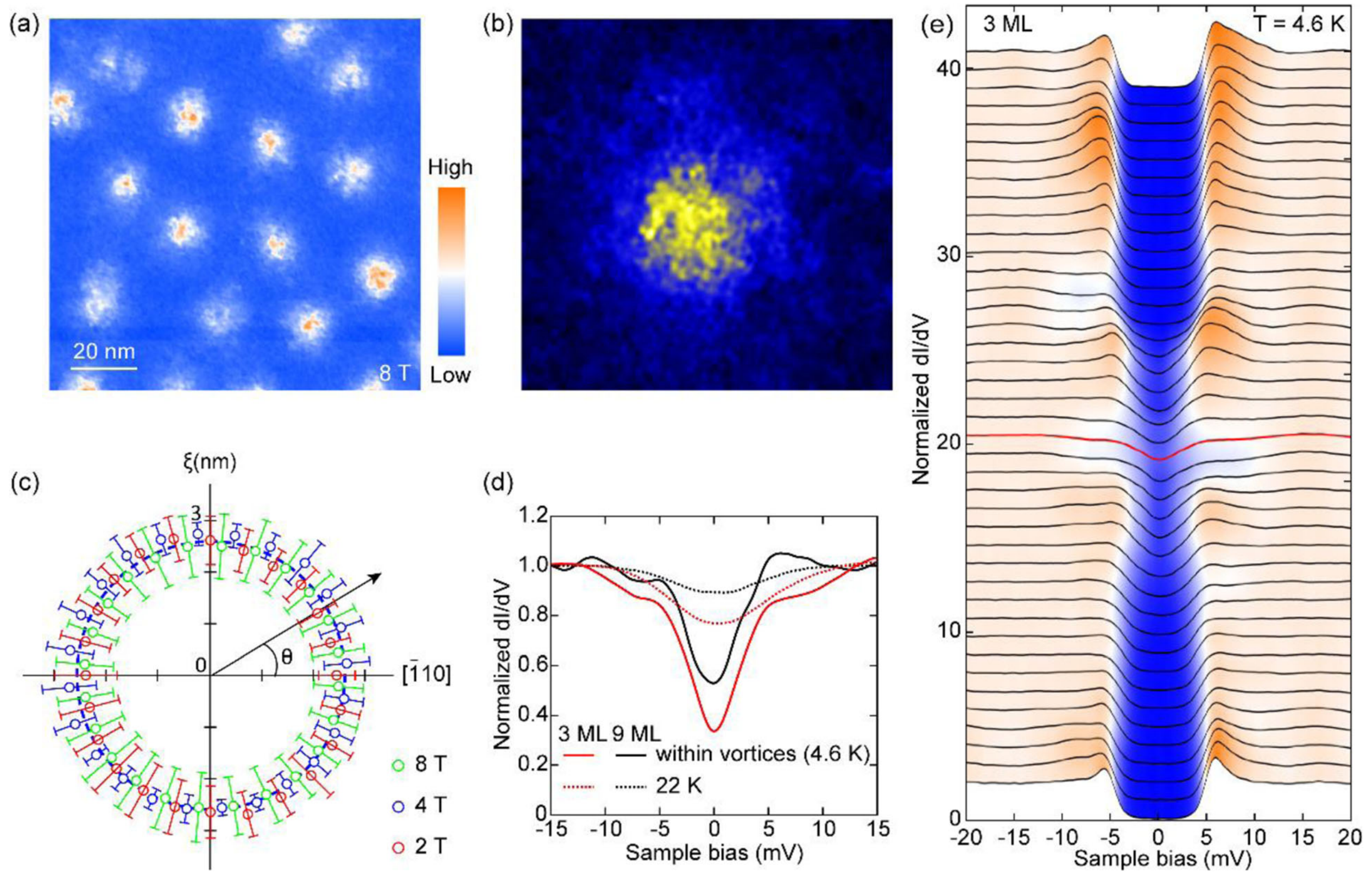

**Fig. 10** Vortex states in 3 ML $K_3C_{60}$. (**a**) Zero energy conductance map under a magnetic field of 8 T. (**b**) Enlarged images for a typical vortex at zero energy. (**c**) Azimuthal dependence of the fitted coherence length ξ. (**d**) Pseudogaps measured at 22 K and within vortices (measured at 4.6 K). (**e**) d*I*/d*V* spectra across a vortex core (*T* = 4.6 K and *B* = 2 T). Taken from ref. [45]

which provide essential insight into the superconducting mechanism. This method has been routinely applied into cuprates and iron-based compounds, but remains unexplored in organic superconductors [74]. It is well documented that non-magnetic impurities little affect the pairing in conventional superconductors [78, 79], but scatter cooper pairs and induce localized in-gap bound states for unconventional pairing symmetry, such as the *d*-wave [79–81] or *s* ± wave superconductor [82–85].

If $A_3C_{60}$ films deviate slightly from stoichiometry, tetrahedral $A^+$ vacancies and octahedra $A^+$ adatoms will appear on the film surfaces, which behave as dark or bright windmills, respectively. They serve as non-magnetic impurities to test the pairing symmetry in fullerides [83]. Figure 11 shows the evolution of superconducting gap across a Rb vacancy (see Figs. 11a, d) and a Rb adatom (see Figs. 11b, e ). No evidence of any bound states is observed on both impurities. The robustness of the superconducting gap structure against alkali-metal atoms is also evidenced in $K_3C_{60}$ [45]. The shrinkage of Δ on Rb adatom arises probably from a local doping variation. Since the alkali metal vacancies are located below the topmost $C_{60}^{3-}$ anions, this renders the local Δ reduction invisible in surface-sensitive STS.

In addition to point defects, step edges can be regarded as one-dimensional perturbations and would bring about Andreev bound states at zero-energy if they are normal to the possible pairing sign-changing direction [86]. These bound states have been observed in a few cuprate and iron-pniticide superconductors [87, 88]. In $A_3C_{60}$, the step edges always run along the close-packed directions of $C_{60}$ molecules, as shown in Fig. 11c. The d*I*/d*V* spectra further reveal a robust superconducting gap near the step edge without any signature of the Andreev bound states (see Fig. 11f).

To further understand the impurity effect in fulleride superconductors, we deposited magnetic Fe atoms on $Rb_3C_{60}$ surface. Fe adatoms occupy the top (Fe-I) and hollow sites (Fe-II) of the $C_{60}$ lattice (see Fig. 12a). The two types of Fe dopants exhibit distinct behaviors (see Fig. 12b). Particularly, a prominent zero-bias conductance peak (ZBCP) is observed on Fe-I. The ZBCP signatures a Yu-Shiba-Rusinov (YSR) state, which is expected from the exchange coupling between the magnetic impurity and an *s*-wave superconductor [78, 79, 81, 83, 89–94]. The coupling strength seems to be dependent on the Fe absorption sites. For Fe-II at the hollow sites, the coupling might be so weak that renders the YSR states nearly merge into the superconducting gap edges and thus are hardly distinguished. Further theoretical analysis is needed to comprehensively understand the Fe registry site-dependent YSR bound states in fulleride superconductors.

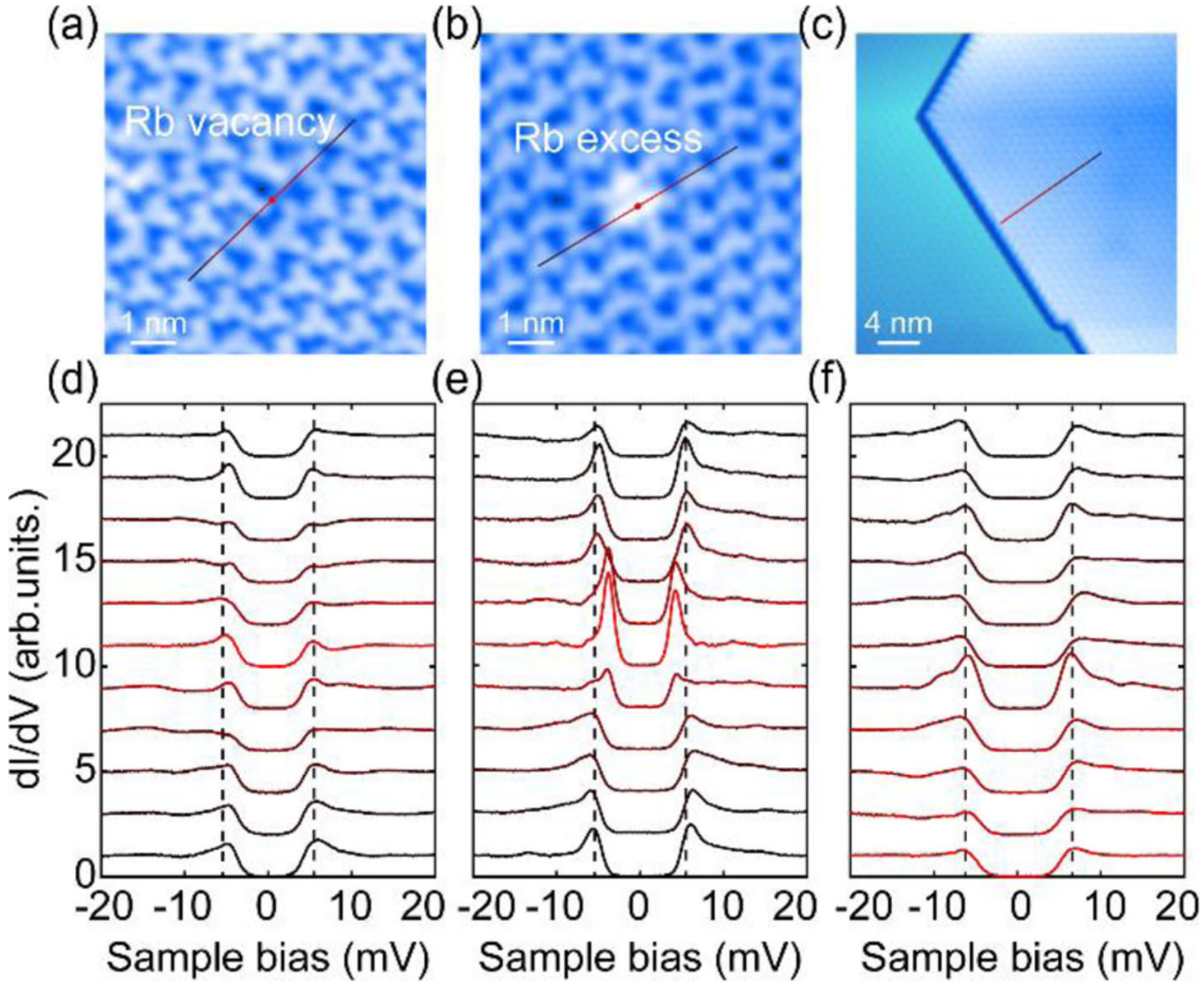

**Fig. 11** (**a**-**c**) STM topographies of a Rb vacancy, Rb excess, and step edge. (**d**-**f**) d*I*/d*V* spectra taken along the colored line in (**a**-**c**), respectively. Taken from ref. [46]

To sum up, the robustness of superconductivity against non-magnetic disorder (alkali-metal atoms vacancies, dopants, and step edges) and emergent YSR bound states on magnetic Fe adatoms, unambiguously supports a sign-unchanged *s*-wave pairing in fulleride superconductors.

### 4.5 Tuning the electronic states and superconductivity via film thickness

For $K_3C_{60}$, the trilayer film processes a higher $T_c$ (~ 22 K) compared to that of 9 ML film (~ 18.4 K) and its bulk counterpart [1, 47]. This indicates that thickness or dimensionality furnishes an alternatively way to tune the electronic properties and superconductivity in $A_3C_{60}$ films. With further reduced film thickness, the superconductivity eventually disappears in $K_3C_{60}$ bilayer and monolayer films. The d$I$/d$V$ spectra on bilayer $K_3C_{60}$ reveal an asymmetric gap ~ 80 meV in the vicinity of $E_F$ (see Fig. 13b). Considering the large gap size, its insensitivity to magnetic field (see Fig. 13c) and the absence of vortices in the zero-energy conduction maps (see Fig. 13d), we conclude that bilayer $K_3C_{60}$ is not superconducting. The DOS hotspots in Fig. 13d, which strongly correlates to excess K dopants, should be assigned as K-induced impurity states. Besides, we find that the DOS intensity near gap edges varies simultaneous with the √3 × √3 superstructure (see Fig. 13b) and the DOS maps exhibit reversal intensity at occupied and empty states (see Fig. 13e, f), suggesting a possible CDW origin for the asymmetric gap in $K_3C_{60}$ bilayer. Moreover, the gap remains robustly in shape and size with elevated temperature up to 40 K (see Fig. 13g), indicating a very high phase transition temperature $T_s$.

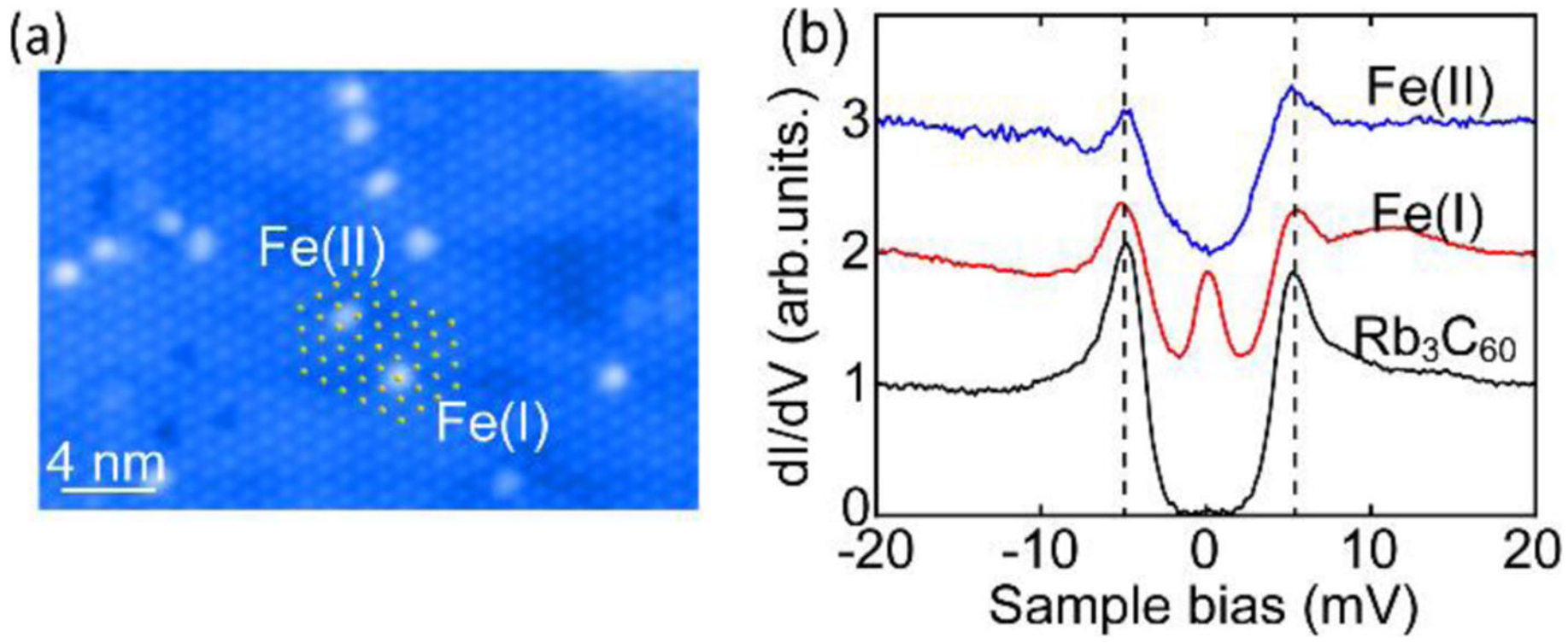

**Fig. 12** Fe-induced bound states. (**a**) STM topography of $Rb_3C_{60}$ films with diluted Fe adatoms. (**b**) Averaged d*I*/d*V* spectra on Fe-I, Fe-II impurities, and locations far from any impurities. Taken from ref. [46]

To reveal the enhanced $T_c$ in trilayer $K_3C_{60}$ and the absence of superconductivity in bilayer films, we explore the layer-dependent structural and electronic properties of $K_3C_{60}$ and depict the d$I$/d$V$ spectra in Fig. 14a. With reduced film thickness, a dip or insulating gap is noticeable near $E_F$ and increases in size, hallmarks of a thickness-controlled metal-insulator transition. The observed tunneling gaps (~ a few hundreds of meV) in monolayer and bilayer $K_3C_{60}$ are too large to be ascribed to CDW correlations. Alternatively, the gaps are suggested to arise from the enhanced electronic correlation $U$ due to the poor screening of $K_3C_{60}$ at the 2D limit [42]. The strong $U$ together with the JT effects splits the $t_{1u}$ orbital and leads to the opening of a Mott gap [44, 45]. Thus, the half-filled monolayer and bilayer $K_3C_{60}$ films become MJTIs at 2D limit. The extracted values of the Hubbard $U$ (energy separation between UHB and LHB) and bandwidth in $K_3C_{60}$ are plotted in Fig. 14b. It is clear that $U/W$ decreases dramatically with reduced film thickness, and the SMIT occurs when $U/W$ ~ 1. The trilayer $K_3C_{60}$ exists on the verge of SMIT and a small enhancement of $U$ renders the bilayer $K_3C_{60}$ not superconducting. In contrast to the SMIT observed in pressurized $Cs_3C_{60}$ and $Rb_xCs_{3-x}C_{60}$ ($0.35 \le x < 2$), which is achieved through a continuous control of $W$ by tuning the interfullerene separation, the SMIT observed here is mainly governed by the thickness-controlled $U$.

The thickness-controlled SMIT is also observed in $Rb_3C_{60}$ films [46], where trilayer $Rb_3C_{60}$ is superconducting, but the increased electronic correlation drives monolayer and bilayer $Rb_3C_{60}$ films into insulating phases. This further proves that the observed SMIT is dominated by thickness-controlled $U$. However, in contrast to the enhanced $T_c$ in trilayer $K_3C_{60}$ film, superconductivity is suppressed in trilayer $Rb_3C_{60}$ (~ 23 K) compared to 9 ML $Rb_3C_{60}$ (~ 28 K) [46]. This is due to a larger molecular spacing in $Rb_3C_{60}$ and the preexisting strong electronic correlation. A further enhancement of *U* pushes thin $Rb_3C_{60}$ films closer to a Mott transitions and thus weakens superconductivity [46].

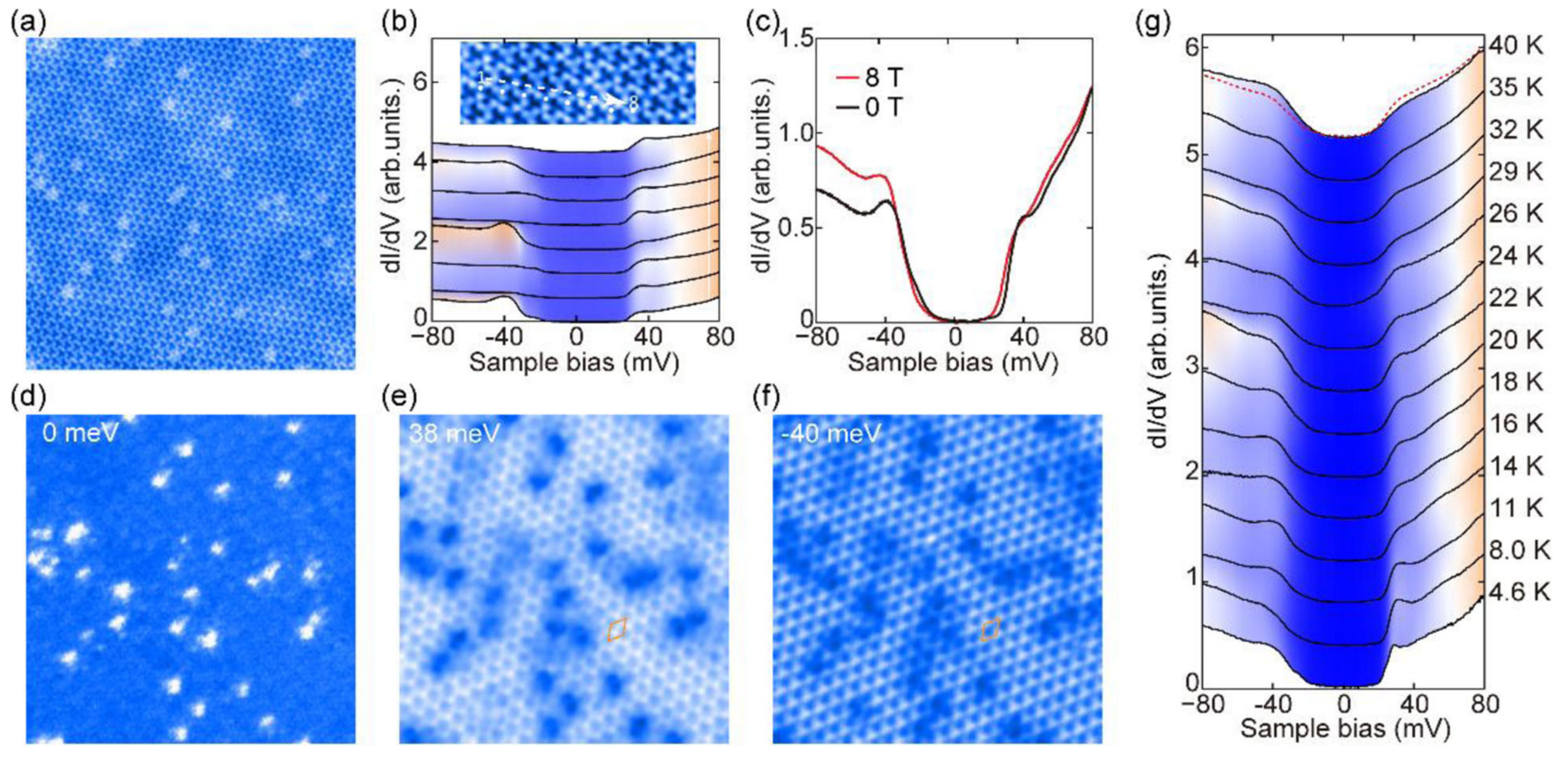

**Fig. 13** Absence of superconductivity in bilayer $K_3C_{60}$ (**a**) STM topography of bilayer $K_3C_{60}$, interspersed with small amount of excess K dopants. (**b**) d$I$/d$V$ spectra along the white dashed line in the inset. (**c**) Averaged d$I$/d$V$ spectra at $B$ = 0 and $B$ = 8 T. (**d**-**f**) Electronic DOS maps with an energy of (**d**) 2 meV, (**e**) 38 meV, and (**f**) −40 meV, acquired simultaneously in the same field with (**a**) at $B$ = 8 T. (**g**) Spatially averaged d$I$/d$V$ spectra as a function of temperature. The red dashed line shows the convolution of electronic DOS measured at 4.6 K with a Fermi-Dirac function with $T$ = 40 K. Taken from ref. [45]

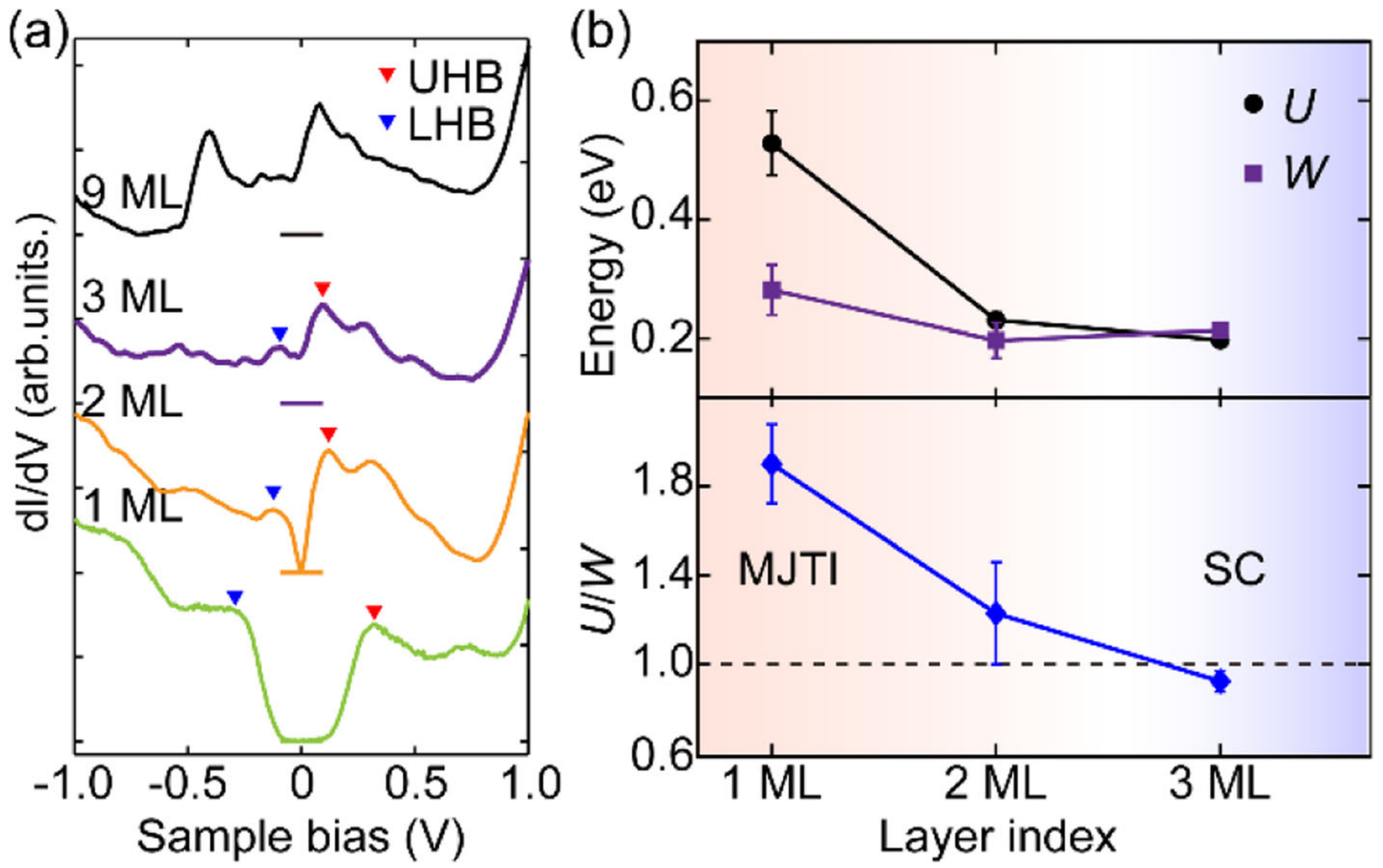

**Fig. 14** (**a**) Thickness-dependent d*I*/d*V* spectra of $K_3C_{60}$. (**b**) Measured Hubbard *U*, *W* (upper panel), and *U*/*W* (lower panel) of $K_3C_{60}$ as a function of thickness. Taken from ref. [45]

The enhanced correlation in $K_3C_{60}$ or $Rb_3C_{60}$ films with reduced thickness is distinct from that of the $K_3C_{60}$ films grown on Au (111) substrate [42]. This discrepancy can be reconciled in consideration of the combined screening effects from nearby polarized $C_{60}{}^{3-}$ anions ($\delta U_p$) and the substrate ($\delta U_s$). On Au (111) substrate, the $\delta U_s$ dominates the Coulomb reduction and results in a smaller effective *U* in the monolayer [42]. On graphene, the substrate screening is dramatically reduced because of the effective decoupling effect as discussed in section 4.1. Alternatively, the larger *U* in monolayer $K_3C_{60}$ or $Rb_3C_{60}$ supported by graphene is attributed to the reduced screening from nearby polarized $C_{60}{}^{3-}$ anions, whose number drops from nine in bilayer to six in monolayer.

Having discussed the electronic properties and superconductivity in stoichiometric $A_3C_{60}$ films, we summarize some of the key parameters, including the Coulomb repulsion (*U*), bandwidth (*W*), transition temperature ($T_c$), superconducting gap size ($\Delta$), reduced gap ratio ($2\Delta/k_BT_c$), and coherence length ($\xi$), of both $K_3C_{60}$ and $Rb_3C_{60}$ thin films in Table 2. The parameters for their bulk counterparts are also listed for comparison. Overall, $A_3C_{60}$ films at single layer limit are MJTIs and evolve into metallic state with reduced correlation or increased film thickness, followed by the emergence of the high-$T_c$ superconductivity, which persists to at least 3 ML films. The $T_c$ and Δ in $A_3C_{60}$ films evolve gradually to that of their bulk counterparts with increased thickness, but deviate significantly in thinner films. Nevertheless, the short coherence length is universal in films and bulks, in accordance to a local pairing picture as will be discussed in section 5.

**Table 2** A brief summary of some parameters of $A_xC_{60}$ films and bulks. Data of thin films are taken from ref. [44–46]

| $K_3C_{60}$ | *U* (eV) | *W* (eV) | $T_c$ (K) | *Δ* (meV) | $2\Delta/k_BT_c$ | *ξ* (nm) |
|---|---|---|---|---|---|---|
| 1 ML | ~ 0.53 | ~ 0.28 | — | — | — | — |
| 2 ML | ~ 0.23 | ~ 0.2 | — | — | — | — |
| 3 ML | ~ 0.2 | ~ 0.21 | 22 | 5.7 | 6.0 | 2.6 |
| 9 ML | — | — | 18.4 | 4.8 | 6.0 | — |
| Bulk | 1.4-1.6 [17, 95] | ~ 0.5 [70] | 18-19 [1, 47] | 4.4 [15] | 3.0-5.3 [15, 16, 96–99] | 2.6-4.5 [75, 100, 101] |
| $Rb_3C_{60}$ | *U* | *W* | $T_c$ | *Δ* (meV) | $2\Delta/k_BT_c$ | *ξ* (nm) |
| 1 ML | ~ 0.84 | ~ 0.30 | — | — | — | — |
| 2 ML | ~ 0.17 | — | — | — | — | — |
| 3 ML | ~ 0.06 | — | 23 | 5.9 | 5.9 | — |
| 6 ML | — | — | 26 | 6.3 | 5.6 | — |
| 9 ML | — | — | 28 | 7.5 | 6.2 | 1.5 |
| Bulk | — | 0.45 [70] | 28-30 [2, 3] | 5.2-6.6 [14, 38, 102, 103] | 3-5.3 [14–16, 38, 56, 97–99, 102, 104, 105] | 2.0-2.4 [100, 103] |

### 4.6 Tuning the electronic states and superconductivity via electron doping

Electron doping provides another way to tune the electronic states of $A_xC_{60}$, because itinerant carriers can significantly screen the on-molecular electronic correlation $U$ [42]. To reveal the doping-controlled $U$, we depict a series of d$I$/d$V$ spectra of $K_xC_{60}$ in Fig. 15. The energy positions of UHB and LHB are marked by the red dashed lines. Apparently, $U$ decreases smoothly with increasing $x$ in all films. This is further evidenced by the fact that the Mott-induced DOS dip at $E_F$ gradually gets shallower with increasing $x$. Taken altogether, a fine control of electronic correlation $U$ via electron doping and thickness has been realized in $K_xC_{60}$ films.

Figure 15a also reveals that bilayer $K_xC_{60}$ undergoes a phase transition from a MJTI at half-filling into a metallic phase with finite DOS($E_F$) at $x$ = 3.36. More interestingly, a low-energy gap of ~ 2.0 meV opens at $E_F$ in $K_{3.36}C_{60}$ bilayer (see Fig. 16c). The amplitude of coherence peaks and gap depth are significantly suppressed under external magnetic fields, confirming a superconductivity origin. Despite somewhat inhomogeneity in coherence peak amplitude, the superconducting gap exists ubiquitously both on and off excess K dopants (see Fig. 16). We therefore realize a doping-controlled SMIT in bilayer $K_xC_{60}$ just like that in cuprate superconductors [74].

In contrast to the monotone shrinkage of $U$, Δ in both 3 ML and 9 ML $K_xC_{60}$ exhibits a dome-shaped variation, with its maximum locked at $x$ = 3 and decreases smoothly when $x$ diverges from 3 (see Fig. 17). Nevertheless, all the d$I$/d$V$ spectra show a fully gapped superconductivity with zero DOS and flat bottoms around $E_F$ [45]. The parabolic-shaped gap in 3 ML $K_{2.928}C_{60}$ results from a limited resolution of our equipment and the spatially averaged effect of the d$I$/d$V$ spectra.

In $K_xC_{60}$ films, we can access a wide range of $x$ (2.7 < × < 3.6) upon K doping. However, phase separation occurs if $x$ derivates dramatically from 3. For $x$ < 3, the K-doped films separate spatially into the superconducting $K_xC_{60}$ with vacancies and pristine $C_{60}$ films. With further K depositing, the $K_xC_{60}$ region enlarges in size and the K vacancies decrease in density, until finally the film evolves into stoichiometric $K_3C_{60}$. The $K_1C_{60}$ and $K_2C_{60}$ do not show up in multilayer films during the entire growth process, indicating that they are less thermodynamically stable than $K_3C_{60}$ on graphene substrate. When $x$ reaches a critical value of ~ 3.3, the K-doped films separate into the superconducting $K_xC_{60}$ ($x$ > 3) and insulating $K_4C_{60}$ [45]. With further K depositing, the superconducting $K_xC_{60}$ gradually decreases in area and evolves into the tetravalent fulleride $K_4C_{60}$ eventually. We also note that the $C_{60}$ molecules in superconducting $K_xC_{60}$ are all H-orientated as that in stoichiometric $K_3C_{60}$, despite of the K dopants or vacancies.

## 5 Phase diagram and superconducting mechanism

In the above section, we have shown the large tunability of the superconductivity and electronic states in $A_3C_{60}$ films by a combined control of alkali-metal atoms, film thickness, and electron doping. This enables us to track for the first time the variations of Δ and $T_c$ over a wide range of electron doping $x$ and thus to plot an unusual phase diagram (see Fig. 18). We note that T. Yildirim et al. have also reported a similar doping-dependent phase diagram of $A_xC_{60}$ crystals [106]. However, the evolution of $T_c$ with doping in their study may be frustrated by the sample diversity and merohedral disorder, which is easily eliminated in our case. Figure 18 reveals several important aspects of the superconductivity in fullerides. (a) In either $K_3C_{60}$ (3 ML, 9 ML) or $Rb_3C_{60}$ (3 ML, 6 ML, and 9 ML), $\Delta(x)$ scales with $T_c$ (lower panel) and exhibits dome-shaped variations with maxima locating at half-filling. (b) Δ declines more rapidly in 9 ML $Rb_xC_{60}$ than in 9 ML $K_xC_{60}$ when $x$ deviates from half-filling. (**c**) In contrast to 9 ML films, $\Delta(x)$ in trilayer fullerides exhibits great asymmetry with respect to $x = 3$. (d) Δ is enhanced in $K_3C_{60}$, but gets suppressed in $Rb_3C_{60}$ films when film thickness reduces from 9 ML to 3 ML.

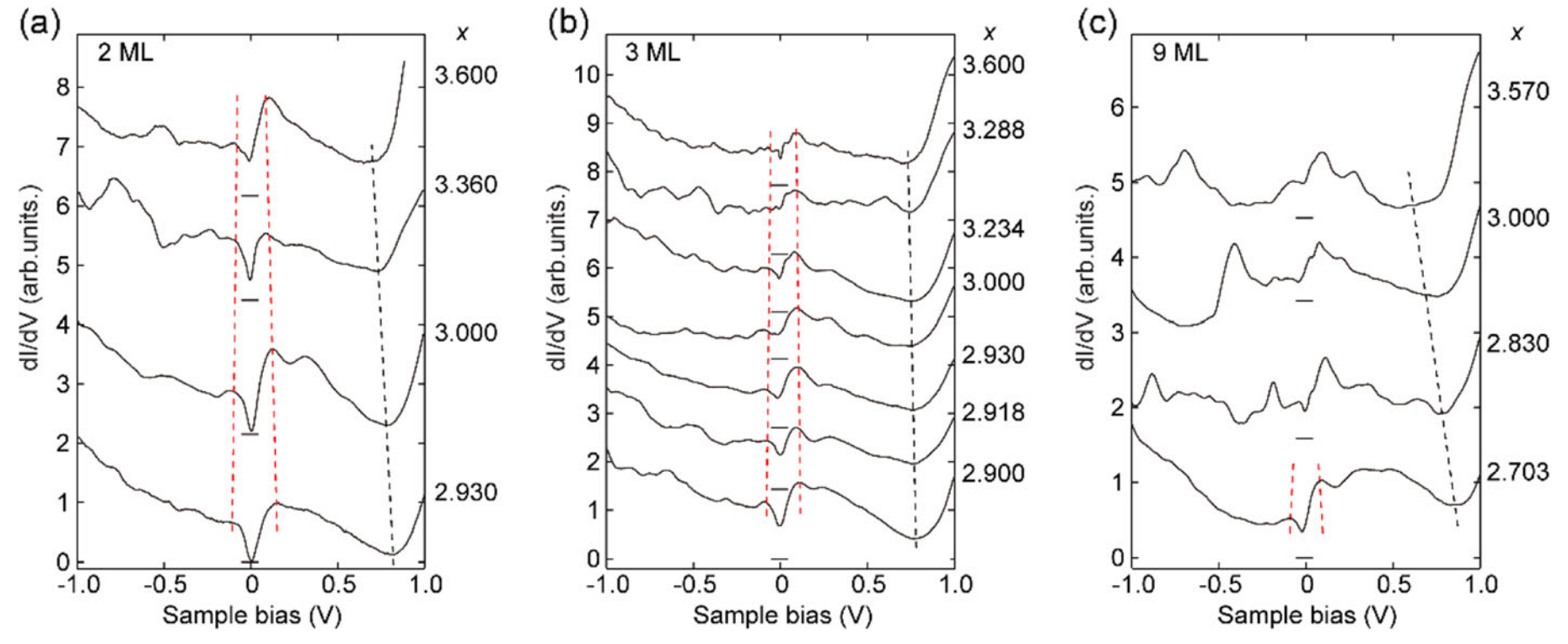

**Fig. 15** Spatially averaged d$I$/d$V$ spectra as a function of electron doping $x$, measured on (**a**) 2 ML; (**b**) 3 ML; (**c**) 9 ML $K_xC_{60}$. The red dashed lines mark the doping evolution of UHB and LHB, with the black dashed ones signifying K-doping. Taken from ref. [45]

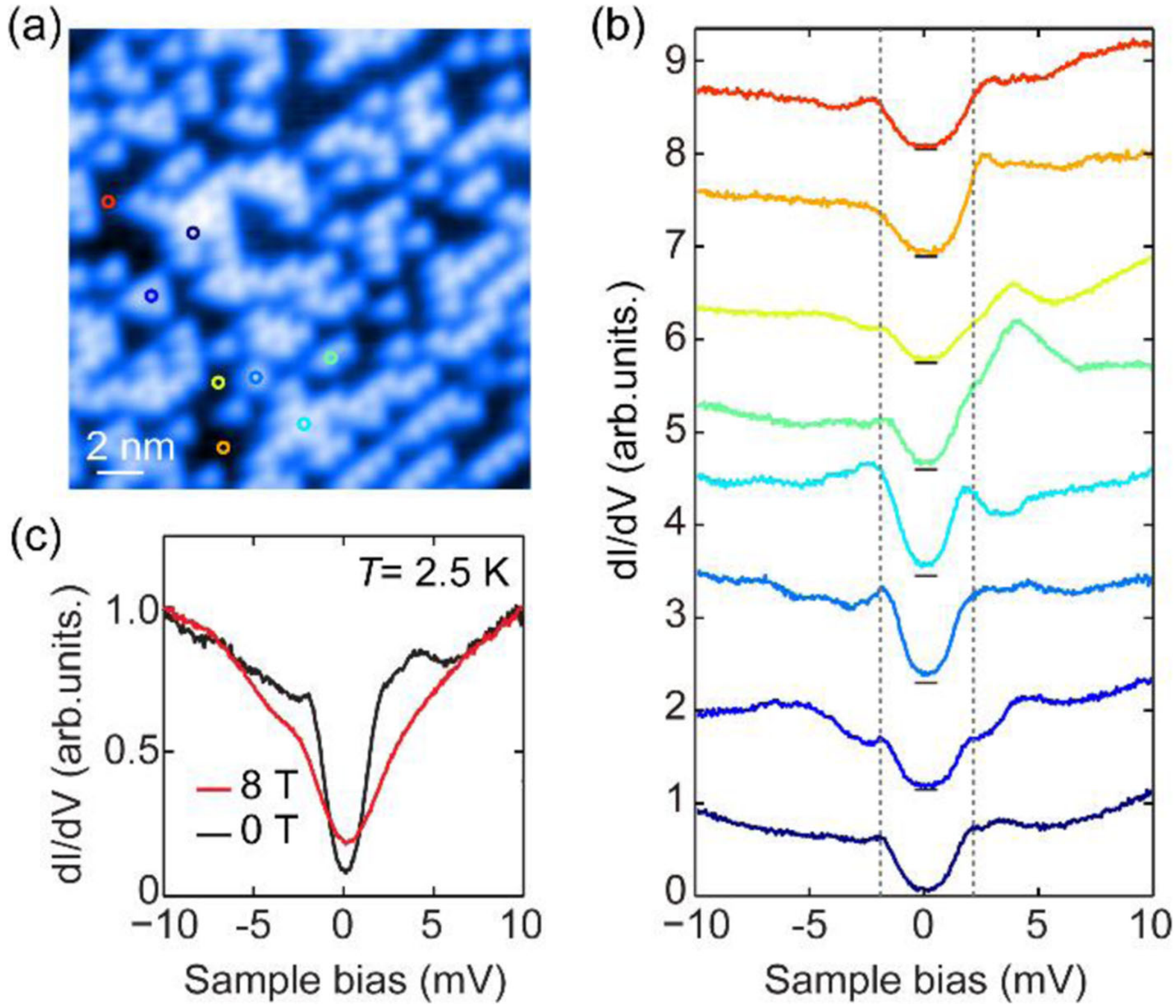

**Fig. 16** Superconductivity in 2 ML $K_{3.36}C_{60}$. (**a**) STM topography of bilayer $K_{3.36}C_{60}$, with excess K dopants randomly distributed. (**b**) d*I*/d*V* spectra measured at $T$ =2.5 K, color-coded to match the probe positions (empty circles) in (**a**). (**c**) Spatially averaged d*I*/d*V* spectra at $B$ = 0 T (black curve) and $B$ = 8 T (red curve). Taken from ref. [45]

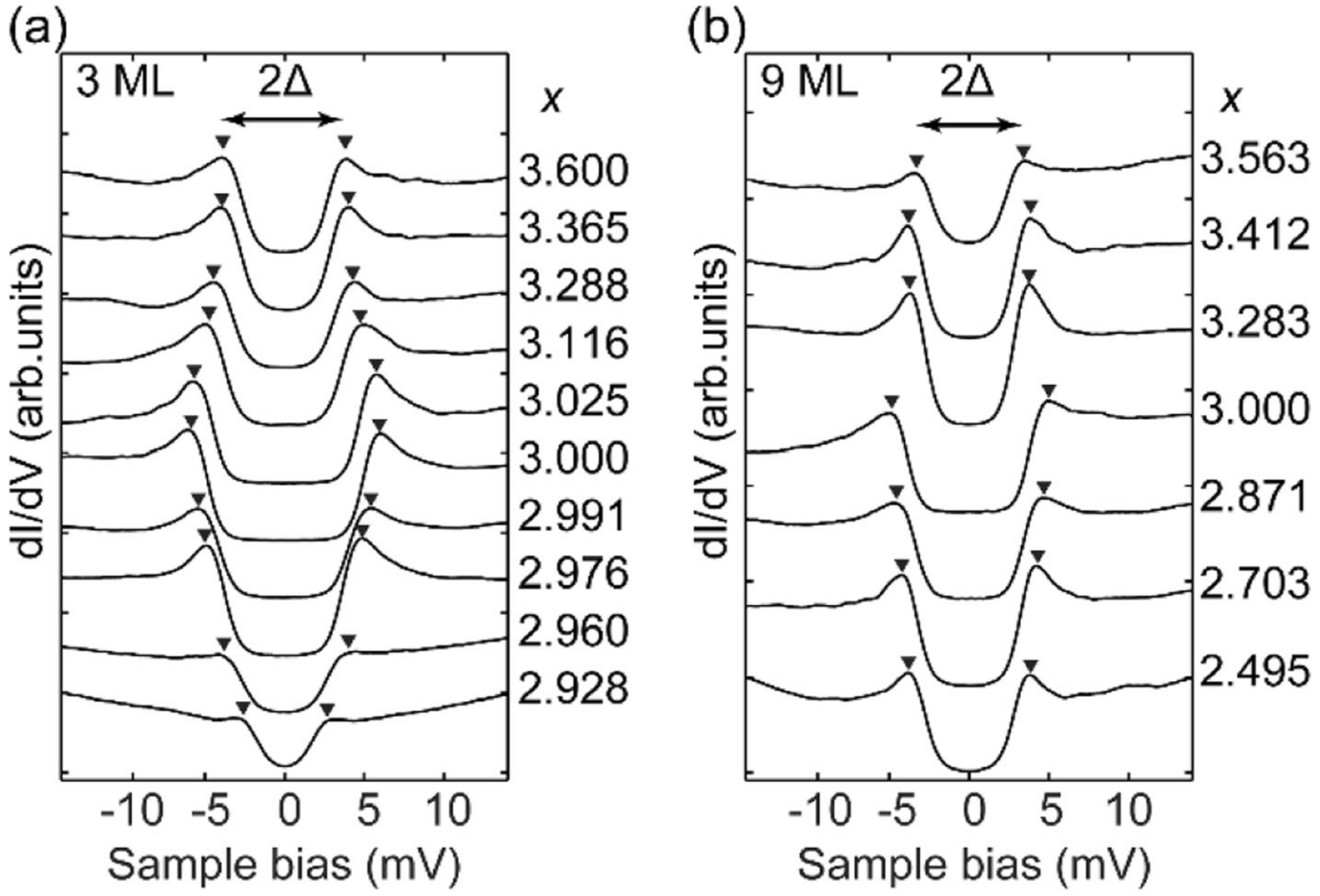

**Fig. 17** Evolution of the averaged superconducting gap (measured at 4.5 K) with K doping $x$ in 3 ML and 9 ML $K_xC_{60}$. The coherence peaks are marked by the black triangles for eye guide. Taken from ref. [45]

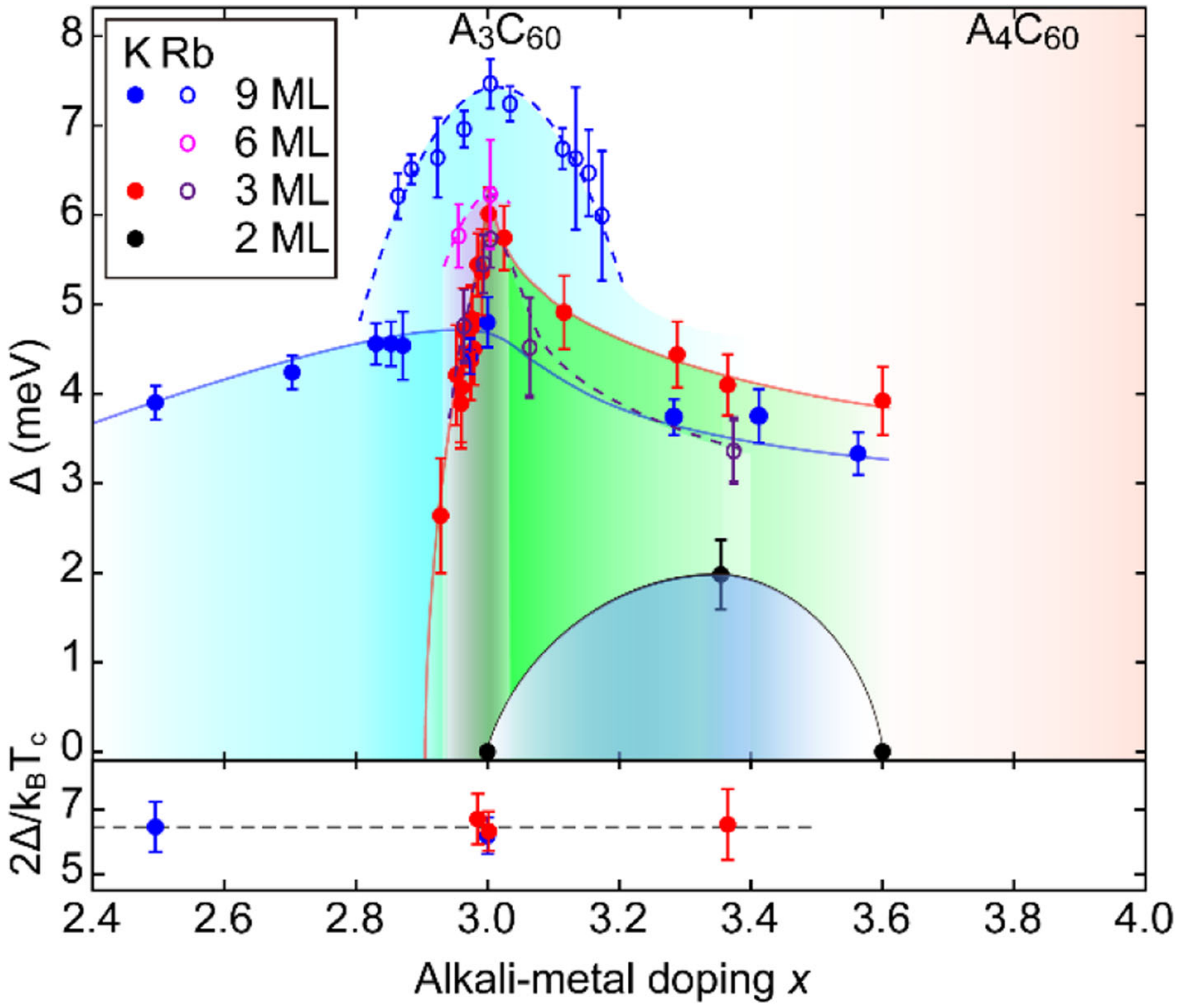

**Fig. 18** Electronic phase diagram showing the evolution of Δ, and reduced gap $2\Delta/k_BT_c$ of $A_3C_{60}$ films as a function of $x$. Taken from ref. [45]

This unique phase diagram imposes strong restrictions on the superconducting mechanism in fullerides. Firstly, the $\Delta(x)$ peaks exactly at half-filling, excluding unambiguously the assumption that superconductivity in $A_xC_{60}$ comes from an accidental doping [17]. Secondly, it seems unlikely that the Δ shrinkage away from half-filling is due to some form of disorder effects, as the $\Delta(x)$ evolution is essentially thickness dependent and of great asymmetry in trilayer fullerides. Particularly, the dome-shaped phase diagram cannot be ascribed to the merohedral disorder effect, because the orientational correlation function remains essentially unchanged with $x$ (see Fig. 4c). Thirdly, the superconducting dome also cannot be attributed solely to the $x$-dependent DOS variation at $E_F$ [32], because the electronic DOS of half-filled $K_3C_{60}$ shows a shoulder and even minimum at $E_F$ (see Fig. 7). Superconductivity would be enhanced with increasing DOS ($E_F$) as $x > 3$, at odds with our observations. The breakdown of the conventional Migdal's theorem might arise from the small retardation in fullerides, since the energy scale of molecular vibrations is comparable to the bare electron bandwidth $W$ [76]. To overcome the weak retardation effects, local pairing is crucial in reducing the effects of Coulomb repulsion.

In the local pairing mechanism, the key ingredient for the high-$T_c$ superconductivity in trivalent fullerides is the strong coupling of the $t_{1u}$ electrons to intramolecular JT phonons. The phonon-mediated multiorbital (attractive) interactions lead to an effectively inverted Hund's coupling ($S = 1/2$) and the formation of a local spin-singlet $s$-wave electron pairs on the same orbital [11, 26, 76]. Superconductivity is further enhanced via a coherent interorbital tunneling of electron pairs (the Suhl-Kondo mechanism) [107, 108]. On the other hand, the multiorbital electronic correlations suppress the electron hopping-induced charge fluctuations and more effectively bind electrons into intraorbital pairs [76]. In this sense, the Coulomb interaction actually helps the local pairing, until it is strong enough to localize the electrons and drive a SMIT.

Such a local pairing picture has also been evidenced by a short coherence length ξ in $A_3C_{60}$ (1.5-2.6 nm), which amounts to only twice the separation between nearest neighboring $C_{60}^{x-}$ [45, 46]. On the other hand, electrons in a superconductor with local non-retarded interactions are paired via a short-range static attraction, and thus the superconductivity is less sensitive to the DOS distribution near $E_F$ [109]. This happens to match the insensitivity of superconductivity in $A_3C_{60}$ to the merohedral disorder, which significantly modifies the $t_{1u}$-derived DOS distribution (see Fig. 7a). The local pairing scenario differs distinctly from a conventional BCS mechanism where only partial electrons near $E_F$ participate in the Cooper pairing and $T_c$ is essentially governed by the DOS($E_F$).

Despite of the dome-shaped $T_c$ that peaks at half-filling, the local pairing scenario also predicts a rapid decay of $T_c$ as $x$ deviates from 3 for a large $U/W$ [76]. This is consistent with our observation that Δ shrinks faster in 9 ML $Rb_xC_{60}$ than in 9 ML $K_xC_{60}$. This scenario naturally explains the enhanced Δ in $K_3C_{60}$ but suppressed Δ in $Rb_3C_{60}$ with reduced film thickness. In $K_3C_{60}$, $U$ is relatively small and its enhancement at reduced film thicknesses stabilizes the local pairing and thus enhances superconductivity, whereas the opposite holds true in $Rb_3C_{60}$ due to the already strong $U$. A further enhancement of $U$ pushes $Rb_3C_{60}$ thin films closer to a Mott transition and thus weakens the superconductivity.

Figure 19 schematically illustrates a unified phase diagram of the charged fullerides. No matter how the electronic correlation $U$ varies with the alkali metal and film thickness, the superconductivity is universally optimal at half-filling. This finding is unusual and most probably stems from a decrease in the dynamical JT-related negative pair binding energy ($U_x$, $x$ is electron doping) away from half-filling [68]. For the evenly charged fullerenes $x$ = 2 or 4, $U_2$ and $U_4$ are positive and the JT coupling instead stabilizes two correlated insulating ground states. Notably, Δ is asymmetric with respect to $x$ = 3 in 3 ML $A_3C_{60}$ thin films as the electronic correlations are strong at 2 D limit. This is related to a monotone shrinkage of $U$ with the doping $x$ in view of the enhanced Coulomb screening from itinerant electron carriers (see Fig. 15). In strongly correlated regimes, a small increase of $U$ below half-filling would result in a Mott localization and leads to the observed dome asymmetry.

As a final remark, we comment on the large gap ratio $2\Delta/k_BT_c$ > 6 in both $K_3C_{60}$ and $Rb_3C_{60}$, which well exceeds the canonical BCS value of 3.53. The large gap ratio seems to be universal in high-$T_c$ narrow-band superconductors, such as iron pnictides, cuprates, and fullerides [45, 56, 58, 74, 83, 85]. The gap large ratio is unexpected from a BCS theory, but falls into the theoretical framework of superconductivity with local nonretarded attractive interactions [109]. Experimentally, a local pairing mechanism has been recently proposed to explain the high-$T_c$ superconductivity in monolayer FeSe epitaxial films on $SrTiO_3$ substrate [51]. A question naturally arises as to whether the local pairing mechanism is applicable to other narrow-band superconductors, such as cuprates and multiband iron pnictides. Another important enlightenment from fulleride superconductors is that the *s*-wave superconductivity is compatible with the strong electronic correlation and the pseudogap. Whether this holds true in cuprates remains unsolved and merits further investigations. Notably, a nodeless pairing is indicated in superconducting copper-oxide monolayer films on $Bi_2Sr_2CaCu_2O_{8+\delta}$ [110].

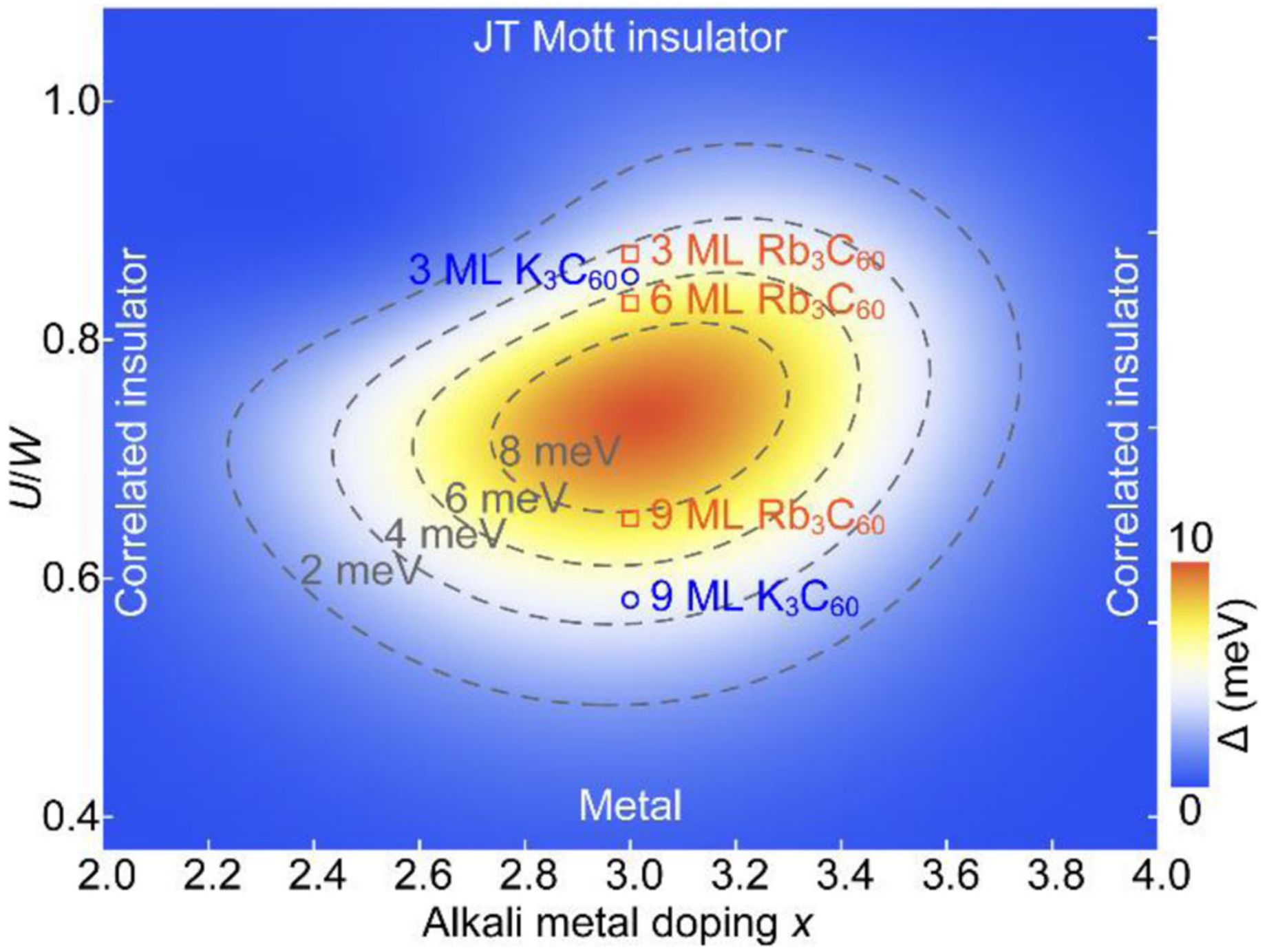

**Fig. 19** Unified phase diagram of Δ variation with electronic correlations $U$ and doping. The empty circles and squares mark the experimental Δ measured in $K_3C_{60}$ and $Rb_3C_{60}$ films, respectively. Contour plots of Δ with a separation of 2 meV are shown in gray dashes. Taken from ref. [46]

Finally, we sum up our recent STM results on the structural, electronic, and superconducting properties of fcc $A_3C_{60}$ films [44–46]. By controlling the alkali-metal species, electron doping, and film thicknesses, $A_xC_{60}$ films exhibit a large tunability of $C_{60}{}^{x-}$ orientations and their orderings. The tunneling spectra of $Cs_xC_{60}$ ($x$ =1, 2, 3, 4) are robustly characteristic of an energy gap, hallmarks of JT instability and strong electronic correlation. With reduced bandwidth or enhanced $U/W$ by substituting Cs with K and Rb, a fully gapped *s*-wave superconductivity accompanied with a cuprate-like pseudogap, emerges in $K_3C_{60}$ and $Rb_3C_{60}$ down to at least 3 ML. The *s*-wave superconductivity exhibits a dome-shaped doping dependence with the optimized gap emerging exactly at half-filling. The strong correlation $U$ drives $\Delta(x)$ to be asymmetric with respect to half-filling in $K_xC_{60}$ and $Rb_xC_{60}$ trilayer films. In any case, the *s*-wave superconductivity retains over the entire phase diagram despite of the electronic correlation and presence of pseudogap. The *s*-wave pairing is further confirmed by its robustness against merohedral disorder, non-magnetic impurities, and step edges. Our experimental results of fulleride superconductors shed important light on the electron pairing in narrow-band high-$T_c$ superconductors.

**Acknowledgements**
The work was financially supported by the China Postdoctoral Science Foundation (2021T140388) and the Natural Science Foundation of China (51788104, 62074092, 11634007).

**Authors' contributions**
M. Q. R. and X. C. M. wrote this manuscript. Review and editing carried out by M. Q. R., C. L. S and Q. K. X. All authors read and approved the final manuscript.

**Funding**
The China Postdoctoral Science Foundation (2021T140388) and the Natural Science Foundation of China.

**Availability of data and materials**
The data that support the findings of this study are available within the article or from the corresponding author upon request.

## Declarations

**Ethics approval and consent to participate**
Not applicable.

**Consent for publication**
Yes.

**Competing interests**
The authors declare that they have no competing interests.

**Author details**
[1]State Key Laboratory of Low-Dimensional Quantum Physics, Department of Physics, Tsinghua University, Beijing 100084, China. [2]Frontier Science Center for Quantum Information, Beijing 100084, China. [3]Beijing Academy of Quantum Information Sciences, Beijing 100193, China. [4]Southern University of Science and Technology, Shenzhen 518055, China.

## Publisher's Note